\newcommand{\CLASSINPUTinnersidemargin}{0.65in}
\newcommand{\CLASSINPUTtoptextmargin}{0.8in}
\newcommand{\CLASSINPUTbottomtextmargin}{1in}

\documentclass[conference]{IEEEtran}
\makeatletter
\def\ps@headings{%
\def\@oddhead{\mbox{}\scriptsize\rightmark \hfil \thepage}%
\def\@evenhead{\scriptsize\thepage \hfil \leftmark\mbox{}}%
\def\@oddfoot{}%
\def\@evenfoot{}}
\makeatother
\usepackage{epsfig,endnotes}

\usepackage{url,graphicx}
\usepackage[cmex10]{amsmath}
\usepackage{enumerate}
\usepackage[ruled,vlined]{algorithm2e}
\usepackage{caption}
\usepackage{subcaption}
\usepackage{color}
\usepackage{balance}
\usepackage{multirow}
\usepackage{adjustbox}

\let\labelindent\relax
\usepackage{enumitem}
\usepackage{cite}

\newcommand{\myparatight}[1]{\smallskip\noindent{\bf {#1}:}}

\newcommand{\sign}{sign}
\newcommand{\name}{\textsc{SybilFuse}\xspace}

\newcommand{\eat}[1]{}

\newcommand{\RG}{\textit{RG}}

\newcommand{\SR}{\textit{SR}}
\newcommand{\CIA}{\textit{CIA}}
\newcommand{\SB}{\textit{SB}}
\newcommand{\SSC}{\textit{SS}}

\newcommand{\SVM}{\textit{SVM}}
\newcommand{\INTEGRO}{\textit{INT}}
\newcommand{\INTEGROPF}{\textit{INT-PF}}
\newcommand{\ENCSB}{\textit{EnC-SB}}
\newcommand{\ENCSR}{\textit{EnC-SR}}
\newcommand{\ENCCIA}{\textit{EnC-CIA}}
\newcommand{\ENCSS}{\textit{EnC-SS}}

\newcommand{\SFLBP}{\textit{SF-LBP}}
\newcommand{\SFRW}{\textit{SF-RW}}

\begin{document}

\title{\name: Combining Local Attributes with Global Structure to Perform Robust Sybil Detection}

\author{
\IEEEauthorblockN{Peng Gao\IEEEauthorrefmark{1},
Binghui Wang\IEEEauthorrefmark{2},
Neil Zhenqiang Gong\IEEEauthorrefmark{2},
Sanjeev R. Kulkarni\IEEEauthorrefmark{1},
Kurt Thomas\IEEEauthorrefmark{3}
and Prateek Mittal\IEEEauthorrefmark{1}}
\IEEEauthorblockA{\IEEEauthorrefmark{1}Princeton University    \IEEEauthorrefmark{2}Iowa State University    \IEEEauthorrefmark{3}Google \\
\IEEEauthorrefmark{1}\{pgao, kulkarni, pmittal\}@princeton.edu    \IEEEauthorrefmark{2}\{binghuiw, neilgong\}@iastate.edu    \IEEEauthorrefmark{3}kurtthomas@google.com
}
}

\maketitle

\begin{abstract}

Sybil attacks are becoming increasingly widespread and pose a significant threat to online social systems; a single adversary can inject multiple colluding identities in the system to compromise security and privacy. Recent
works have leveraged social network-based trust relationships to defend against Sybil attacks. However, existing defenses are based on oversimplified assumptions about network structure, which do not necessarily hold in real-world social networks. 
Recognizing these limitations, we propose \name, a defense-in-depth framework for Sybil detection when the oversimplified assumptions are relaxed. \name adopts a collective classification approach by first training local classifiers to compute local trust scores for nodes and edges, and then propagating the local scores through the global network structure via 
weighted random walk and loopy belief propagation mechanisms. 
We evaluate our framework on both synthetic and real-world network topologies, including a large-scale, labeled Twitter network comprising 20M nodes and 265M edges, and demonstrate that \name outperforms state-of-the-art approaches significantly. In particular, \name achieves 98\% of Sybil coverage among top-ranked nodes. 

\end{abstract}

\IEEEpeerreviewmaketitle

\section{Introduction}
\label{sec:intro}

Our systems today are vulnerable to Sybil attacks, in which an attacker injects multiple fake accounts into the system~\cite{Sybil}.
Recently, the increasing popularity of online social networks has made them attractive targets for Sybil attacks. It is estimated that tens of millions of Sybil accounts exist in popular social networks such as Facebook and Twitter~\cite{Facebooksybil, benevenuto2010detecting}. 
The attacker can leverage Sybil accounts to disrupt democratic election and influence financial market via spreading fake news~\cite{election,stock}, 
as well as compromise system security and privacy via propagating social malware, disseminating scams, 
and learning users' private data~\cite{Facebooksybil,benevenuto2010detecting,Thomas11-imc,Thomas11,Bilge09,Fong11,thomas:leet12}.

\myparatight{Limitations in existing Sybil defenses}
An important thread of research proposes to mitigate Sybil attacks using social network-based trust relationships.
The key insight is that in a social graph where edges represent strong trust relationships, it is hard for the attacker to set up links to benign users. As a result, the number of edges between benign users and Sybil identities (called \emph{attack edges}) is limited.
Approaches such as SybilGuard~\cite{Yu06}, SybilLimit~\cite{Yu08}, SybilInfer~\cite{Danezis09}, SybilRank~\cite{sybilrank}, CIA~\cite{Yang12-spam}, SybilBelief~\cite{Gong13}, and SybilSCAR~\cite{wang2017sybilscar} rely on such \emph{strong trust assumption} and separate the benign and Sybil regions by identifying communities~\cite{Viswanath10}. \'{I}ntegro~\cite{integro} extends SybilRank by considering victim prediction (i.e., benign accounts that connect to Sybils). %

While these approaches have pioneered the use of social network structure for Sybil defenses, the actual deployment of these ideas in real-world networks remains controversial. 
Real-world social networks do not necessarily have strong trusts. Yang et al. showed that RenRen, the largest social network in China, does not follow this assumption~\cite{Yang11-sybil}. Previous work~\cite{Bilge09, boshmaf:acsac11} also showed that Sybil nodes could befriend benign users on Facebook at a large scale. Ghosh et al.~\cite{Ghosh12} showed that on Twitter, a link farming phenomenon is widespread, in which certain benign accounts blindly follow back accounts who follow them. On such weak trust 
networks, 
the number of attack edges can be larger than what is typically considered in previous works, making it challenging to separate the benign region from the Sybil region.
\myparatight{\name outperforms state-of-the-art}
Complex attack strategies in real-world social networks make it hard for methods that are based on single source of information to succeed. 
Recognizing the limitations in existing approaches, we propose \emph{\name}, a defense-in-depth framework that leverages heterogeneous sources of information to perform robust Sybil detection. 
Different from existing approaches which assume strong trust networks~\cite{Yu06,Yu08,Danezis09,sybilrank,Yang12-spam,Gong13,jia2017random,wang2017sybilscar} or assume strong victim prediction~\cite{integro}, \name relaxes these limitations by adopting a collective classification scheme. 
Given social network data as input, \name first leverages \emph{local attributes} to train local classifiers. Local node classifier predicts a trust score for each node, which indicates the probability of that node to be benign. Local edge classifier predicts a trust score for each edge, which indicates the probability of that edge to be a non-attack edge. These local trust scores are then combined with the \emph{global structure} through weighted trust score propagation. Existing approaches do not leverage rich local information and treat edges equally, thus do not work well when the number of attack edges exceeds their assumption.
In contrast, \name captures local account information in \emph{node trust scores}, and propagates these scores through the global structure. During the score propagation, \name leverages \emph{edge trust scores} to enforce unequal weights, so that attack edges will have reduced impacts on the propagation. After the propagation completes, final trust scores of accounts are used for Sybil classification and ranking.

\myparatight{Evaluation}
We conduct comprehensive evaluations of \name against state-of-the-art approaches:
(1) we extensively evaluate the robustness of \name under different network settings and observe that \name is robust to errors in local classifiers (Section~\ref{subsec:sync_res});
(2) we evaluate \name in a real-world, labeled Twitter network in which the assumptions that existing approaches
require are not satisfied (e.g., large number of attack edges (49.5 per Sybil) and victims (75.4\% of benign nodes)). We observe that \name outperforms state-of-the-art approaches significantly in Sybil ranking and achieves 98\% of Sybil coverage among top-ranked nodes (Section~\ref{subsec:small_twitter_results});
(3) we evaluate \name in a real-world, large-scale, labeled Twitter network comprising over 20 million nodes and 265 million edges. We observe that \name outperforms state-of-the-art approaches significantly in both AUC and Sybil ranking (Section~\ref{subsec:large_twitter_results}). 
\eat{
We validated our findings with Twitter's security team and find that \name is able to capture a large portion of Sybil accounts that Twitter fails to detect (Section~\ref{subsubsec:top_100}).
}

\section{Background and Related Work}
\label{sec:background}
\subsection{Sybil Attack Scenario}
Consider a network topology $G = (V, E)$, comprising a set $V$ of nodes (i.e., user accounts) and a set $E$ of edges (i.e., friendship relationships). To model real-world social networks, graph $G$ can be either directed or undirected. A directed graph models the follower-following topology (e.g., Twitter), in which $(u,v)\in E$ denotes that $u$ follows $v$. An undirected graph models the mutual relationship topology (e.g., Facebook), in which $(u, v)\in E$ is equivalent to $(v,u)\in E$.

\begin{figure}[t]
\centering
\includegraphics[width=0.24\textwidth]{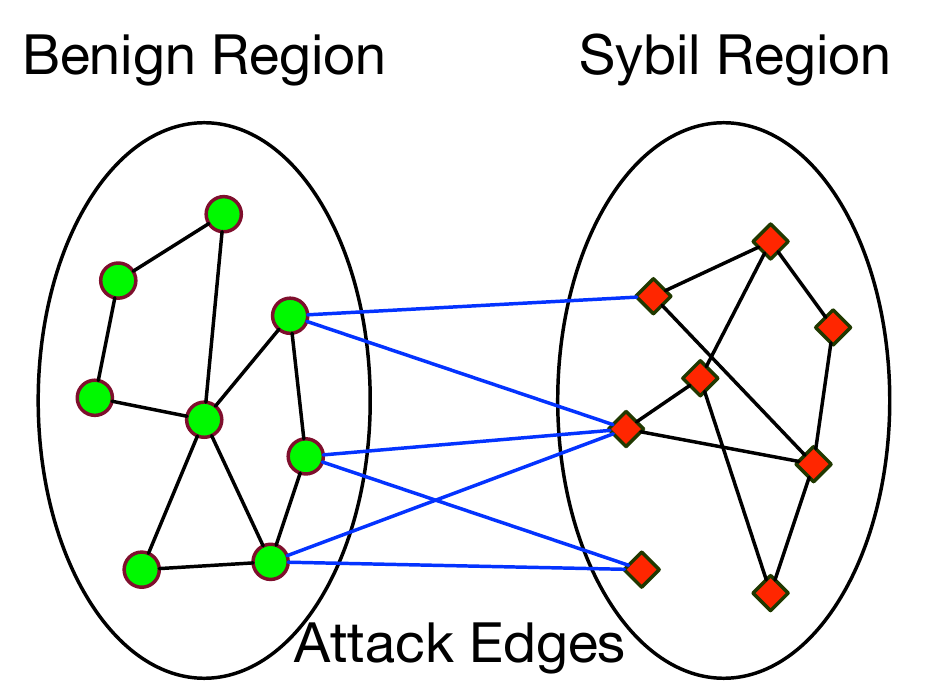}
\caption{Sybil attack scenario}
\label{fig:sybil_attack}
\vspace*{-4ex}
\end{figure}

Fig.~\ref{fig:sybil_attack} shows the \emph{Sybil attack} scenario, 
in which every node $v \in V$ is associated with a label that indicates its identity to be \emph{benign} or \emph{Sybil}. We denote the subnetwork containing all benign nodes to be the benign region, and denote the subnetwork containing all Sybil nodes to be the Sybil region. The edges that connect the two regions are called \emph{attack edges}. 
Following the established convention in the literature, we do not impose any constraints on the size or the shape of the Sybil region. The attacker can create an unlimited number of Sybil nodes and set up edges between them arbitrarily. The main goal of Sybil defense research is to design a mechanism to detect as many Sybil nodes as possible, while minimizing the number of benign nodes that are misdetected.

\subsection{Limitations in State-of-the-art Sybil Defenses}
\label{subsec:approaches}

\myparatight{Local attribute-based approaches}
Local attribute-based approaches seek to detect Sybil accounts by analyzing local account attributes (e.g., posts, status updates, connections).
These approaches span a large category of schemes, including blacklisting~\cite{Ramachandran07}, whitelisting~\cite{yardi10}, URL filtering~\cite{Thomas11}, as well as various machine learning methods, such as Bayesian reasoning, Support Vector Machines, and clustering~\cite{Wang10, gao2012towards, spam:acsac10}.
A fundamental limitation in these approaches is that Sybils can easily evade the detection by mimicking the local behaviors of benign users via manipulating their profiles and connections.

\myparatight{Global structure-based approaches}
Recognizing the limitations in local attribute-based approaches, global structure-based approaches seek to exploit the global graph-theoretic differences between the benign region and the Sybil region.
Most global structure-based approaches leverage either \emph{random walks} or \emph{loopy belief propagation}. For instance, random walk based approaches include SybilGuard~\cite{Yu06}, SybilLimit~\cite{Yu08}, SybilInfer~\cite{Danezis09}, SybilRank~\cite{sybilrank}, CIA~\cite{Yang12-spam}, and SybilWalk~\cite{jia2017random}. \'{I}ntegro~\cite{integro} extends SybilRank by incorporating victim prediction (i.e., benign accounts that connect to Sybils) in random walks. 
Loopy belief propagation based approaches include SybilBelief~\cite{Gong13} and GANG~\cite{wang2017gang}. Fu et al.~\cite{robustspammer} extended SybilBelief via considering user carefulness at establishing social relationships. Wang et al.~\cite{wang2017sybilscar} proposed a local rule based framework  to unify random walk based and loopy belief propagation based approaches. Based on the framework, they further proposed SybilSCAR, which can be viewed as an optimized version of SybilBelief.

Research has shown that~\cite{Yu06,Yu08,Danezis09,sybilrank,Yang12-spam,Gong13,jia2017random,wang2017sybilscar} assume  a \emph{strong trust network}, where the number of attack edges is limited~\cite{Viswanath10}. 
\cite{Yu06,Yu08} further assume that the benign region is fast mixing~\cite{mohaisen:imc10}, which presumes the existence of a well-connected, giant community structure of benign users. 
However, these assumptions oversimplify the network structure and may not hold well on certain real-world social networks.
First, real-world social networks do not necessarily have strong trusts. Yang et al. showed that RenRen, the largest social network in China, does not follow this assumption~\cite{Yang11-sybil}. Previous work~\cite{Bilge09, boshmaf:acsac11} also showed that Sybil nodes could befriend benign users on Facebook at a large scale. Ghosh et al.~\cite{Ghosh12} showed that on Twitter, a link farming phenomenon is widespread, in which certain benign accounts blindly follow back accounts who follow them. On such weak trust 
networks, the number of attack edges can be larger than what is typically considered in previous works, making it challenging to separate the benign region from the Sybil region.
Second, benign users tend to form multiple small communities~\cite{Viswanath10} driven by different purposes (e.g., geographical location, education, and career), which introduces a longer mixing time than the theoretical anticipated value~\cite{mohaisen:imc10}.
Third, although \'{I}ntegro does not directly require a small number of attack edges,
it relies on strong assumptions that the number of victims is small and that the victims are accurately predicted, which may not hold on real-world social networks like Twitter (Section~\ref{subsec:small_twitter_data}).

\begin{figure*}[t]
\centering
\includegraphics[width=0.8\textwidth]{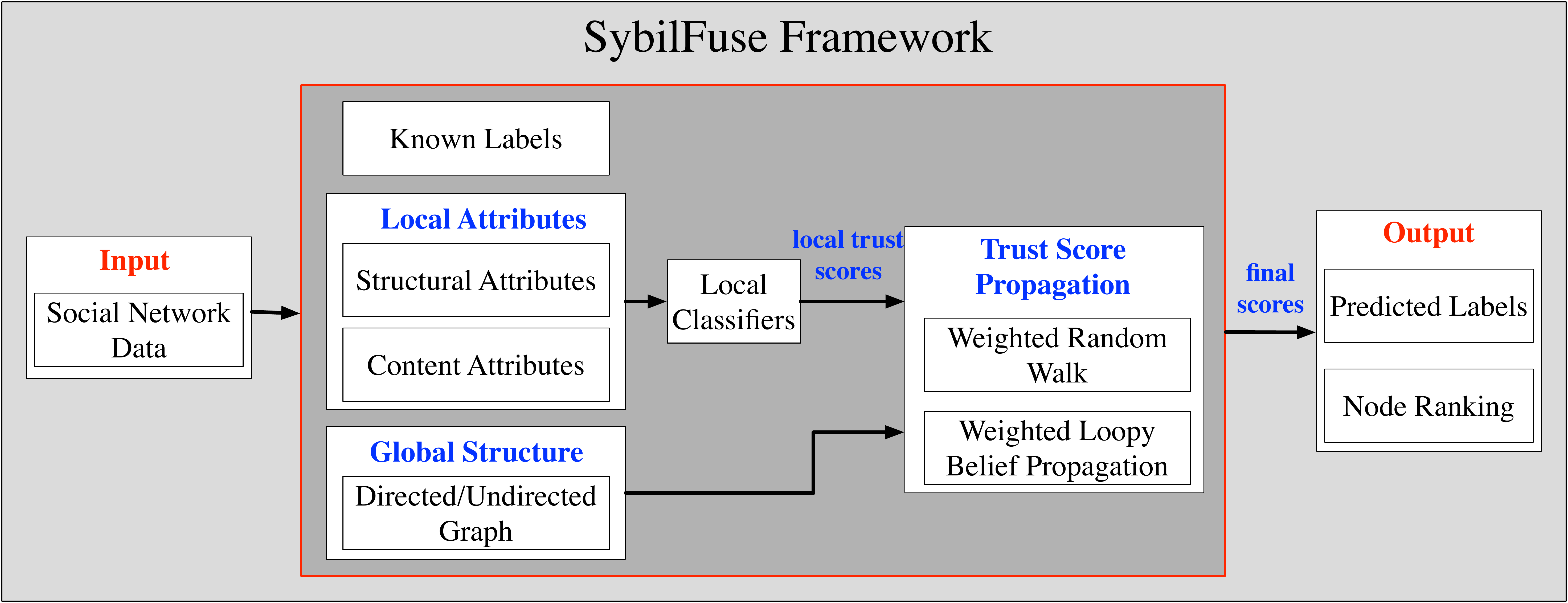}
\caption{The \name framework} 
\label{fig:sybilframe}
\vspace*{-3ex}
\end{figure*}

\section{The \name Framework}
\label{sec:framework}

Complex attack strategies in real-world social networks make it hard for methods that are based on single source of information to succeed. Recognizing the limitations in existing approaches, we propose \emph{\name}, a defense-in-depth framework that leverages heterogeneous sources of information to perform robust Sybil detection. Different from existing approaches which only leverage local attribute information~\cite{Ramachandran07,yardi10,Thomas11,Wang10, gao2012towards, spam:acsac10}, only leverage global structure information~\cite{Yu06,Yu08,Danezis09,sybilrank,Yang12-spam,Gong13,jia2017random,wang2017sybilscar}, or assume strong victim prediction~\cite{integro}, \name combines local attributes with global structure by adopting a collective classification scheme.

\subsection{Framework Overview}
\label{subsec:overview}

Fig.~\ref{fig:sybilframe} shows the \name framework.
Given social network data as input, \name first samples training data and leverages discriminative local attributes (structural or content) to train local classifiers. Local node classifier predicts a trust score for each node, which indicates the probability of that node to be benign. Local edge classifier predicts a trust score for each edge, which indicates the probability of that edge to be a non-attack edge.
These local 
scores are then combined with the global structure through weighted trust score propagation. 
Existing approaches~\cite{Yu06,Yu08,Danezis09,sybilrank,Yang12-spam,Gong13} do not leverage rich local information (i.e., these approaches propagate manually-set scores from a limited set of labeled seeds) and treat edges equally, thus do not work well when the number of attack edges exceeds their assumptions.
In contrast, \name captures local account information in node trust scores, and propagates these scores through the global structure. 
\name furthermore leverages edge trust scores to enforce unequal weights for the propagation, so that attack edges will have reduced impacts. Final node scores after propagation are used for Sybil account prediction and ranking.

\subsection{Local Trust Score Computation}
\label{subsec:prior}

Given a social graph $G = (V, E)$, we denote the trust score of node $v \in V$ by $S_v$, which quantifies the probability that $v$ is benign. We denote the trust score of edge $(u, v)\in E$ by $S_{u, v}$, which quantifies the probability that node $u$ and node $v$ take the same label (i.e., homophily strength). 
To compute $S_v$, we leverage local node attributes (e.g., degree, local clustering coefficient, profile) and train a machine learning classifier that outputs probabilistic estimates (e.g., SVM, logistic regression). 
To compute $S_{u, v}$, we similarly build an edge classifier. In addition, we can measure the similarity between node $u$ and $v$ using various metrics (e.g., Cosine, Jaccard, Adamic-Adar~\cite{Liben-Nowell2007}). The insight is that in social networks where homophily exists, connected benign nodes tend to be similar and Sybils might be different from their targeted benign nodes. As a result, attack edges tend to have low similarity scores.
Note that in practice, we restrict $S_v$ and $S_{u, v}$ to be within range $[0.1, 0.9]$ through normalization, since scoring zero would invalidate our weighted score propagation.

\subsection{Weighted Trust Score Propagation}
\label{subsec:score_propagation}

Score propagation is done through either weighted random walk or weighted loopy belief propagation.

\myparatight{Weighted random walk} 
Let $S^{(i)}(v)$ denote the score of node $v$ after $i$-th power iteration.
First, we set the initial score of every node (excluding the training data) to be the predicted local node trust score from the local node classifier:
\begin{equation}
\label{eq:sr_init}
	S^{(0)}(v) = S_v
\end{equation}
For nodes that belong to the training data, we set score 0.9 to the benign nodes and score 0.1 to the Sybil nodes.

Next, we set the weight of every edge to be the predicted local edge trust score from the local edge classifier, and start weighted random walk using the following update equation:
\begin{equation}
\label{eq:sr_prop}
S^{(i)}(v) = \sum_{(u, v) \in E}S^{(i-1)}(u)\frac{S_{u, v}}{\sum_{(u, w) \in E}S_{u, w}}
\end{equation}
Hence, $u$ will distribute more of its round $i-1$ trust to a close friend $v$ (i.e., $S_{u, v}$ is high) rather than an unfamiliar connection $w$. In this way, attack edges that have low trust scores will have reduced impacts on the propagation.
After $d = O(\log n)$ steps of power iteration (i.e., early termination), where $n$ is the number of nodes, we obtain the final score:
\begin{equation}
S^F_v = S^{(d)} (v)
\end{equation}

\myparatight{Weighted loopy belief propagation}
We model the graph $G$ as a pairwise Markov Random Field (MRF)~\cite{Cross83}. For each node $v$, we associate it with a binary random variable $X_v \in \{1, -1\}$ that represents its label ($X_v = 1$ for benign and $X_v = -1$ for Sybil).  To quantify the probability of a collection of nodes jointly taking a set of labels, we introduce \emph{node potential function} $\psi_v(X_v)$ for node $v$ and \emph{edge potential function} $\psi_{u, v}(X_u, X_v)$ for edge $(u, v)$, and initialize them using our predicted trust scores from local classifiers:
{\footnotesize
\begin{equation}
  \psi_v(X_v)=\begin{cases}
  	S_v & \text{if }X_v = 1 \\
  	1 - S_v & \text{if }X_v = -1
  \end{cases}  
\end{equation}
\begin{equation}
	\psi_{u, v}(X_u, X_v)=\begin{cases}
		S_{u, v} & \text{if }X_uX_v = 1 \\
		1 - S_{u, v} & \text{if }X_uX_v = -1
	\end{cases}  
\end{equation}
}%
Given a pairwise MRF $(G, \Psi)$, we propagate local trust scores through the global structure using Loopy Belief Propagation (LBP)~\cite{Murphy99} with the following message update function:
{\footnotesize
\begin{align}
\begin{split}
&m_{u\rightarrow v} (X_v) \\= &\sum_{X_u}\left( \psi_u(X_u)\psi_{u, v}(X_u, X_v)\prod_{s\in Neighbors(u)\backslash v}m_{s \rightarrow u}(X_s) \right)
\end{split}
\end{align}
}%
where $m_{u\rightarrow v} (X_v)$ is initially set to 1 for all edges $u\rightarrow v$.
Note that edge potentials enforce unequal contribution of edges to the propagation.
After $d = 5\sim 10$ rounds of iteration (we validated this empirically), we obtain the belief score of node $v$ with label $X_v = x_v$:
{\footnotesize
\begin{equation}
bel_v(X_v = x_v) \propto \psi_v(X_v = x_v)\prod_{u\in Neighbors(v)}m_{u\rightarrow v}(X_v = x_v)
\end{equation}
}%
We then normalize $bel_v(X_v = 1)$ to obtain the final score:
{\footnotesize
\begin{equation}
S^F_v = \frac{bel_v(X_v = 1)}{bel_v(X_v = 1) + bel_v(X_v = -1)}
\end{equation}
}%

\myparatight{Computational complexity}
The complexity of weighted random walk and weighted LBP is both $O(md)$, where $m$ is the number of edges and $d$ is the number of iterations (recall that $d = O(\log n)$ for weighted random walk and $d = 5\sim 10$ for weighted LBP). For sparse 
networks, $O(md) = O(nd)$. Thus, the total computational cost is the addition of execution time of local classifier training and prediction, and that of random walk/LBP. In practice, local classifier can be trained offline and efficient implementations like \emph{LIBSVM}~\cite{Chang2011} exist. Besides, random walk and LBP are easily parallelizable, which further speeds up the execution.

\subsection{Sybil Account Prediction and Ranking}
\label{subsec:classify}

\myparatight{Sybil account prediction}  
For a node $v$, we predict its label $L_v$ by comparing its final score $S^F_v$ with a threshold value:
\begin{equation}
L_v = \sign(S^F_v - threshold)
\end{equation}
where $L_v = 1$ indicates a benign label and $L_v = -1$ indicates a Sybil label. The threshold value can be obtained by conducting cross-validation on the training data.

\myparatight{Sybil ranking}
We can also surface Sybil accounts by ranking all nodes in an ascending order of their final scores. Sybils with low scores will be ranked upfront. Security analysts can then go through the list and surface Sybil accounts manually.

\begin{figure*}[!h]
\center
\begin{subfigure}[H]{0.27\textwidth}
  \includegraphics[width=\linewidth]{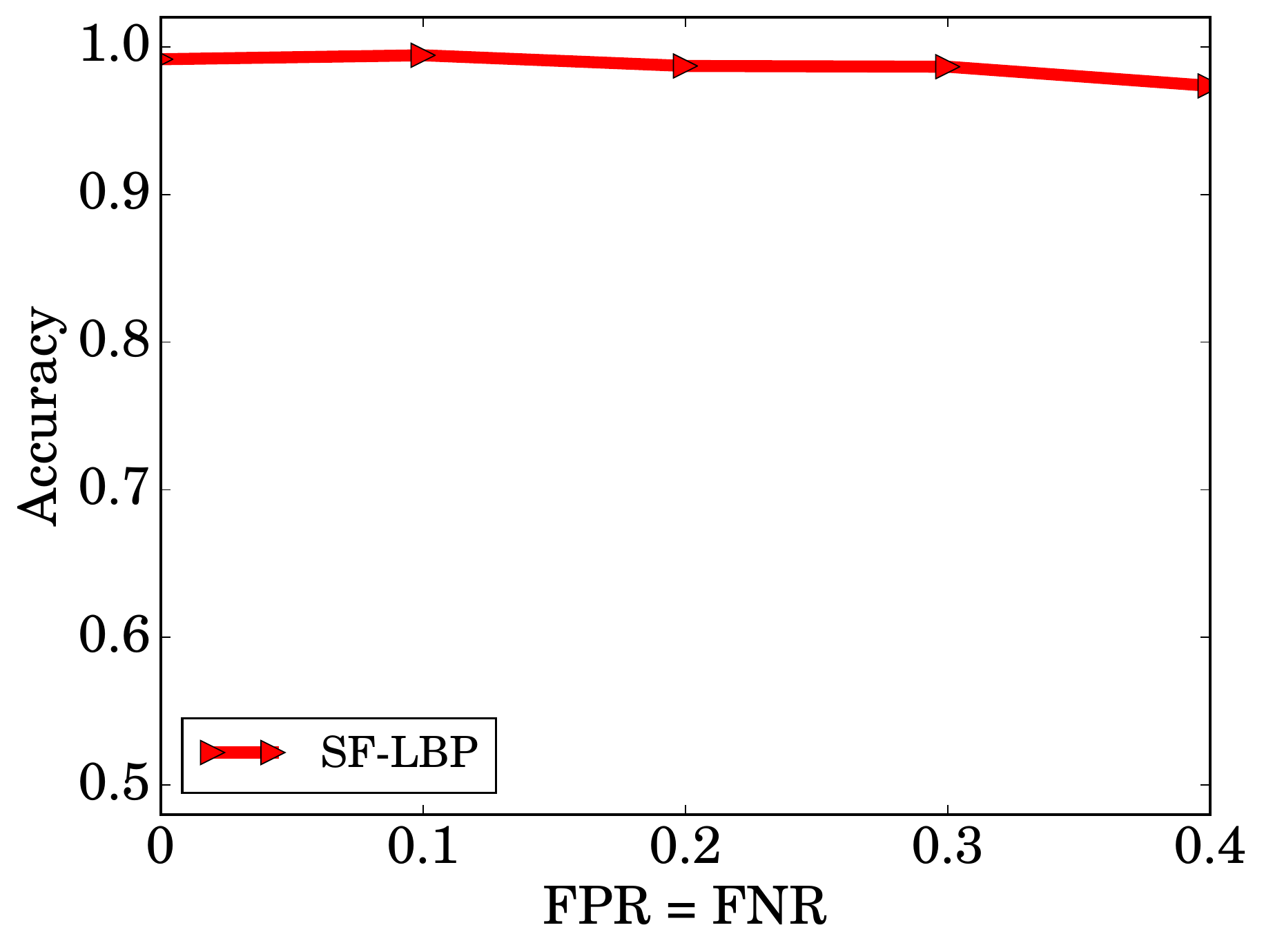}
  \caption{Vary FPR=FNR}
  \label{fig:node_prior_accuracy_1}
\end{subfigure}%
\hspace{0.5cm}
\begin{subfigure}[H]{0.27\textwidth}
  \includegraphics[width=\linewidth]{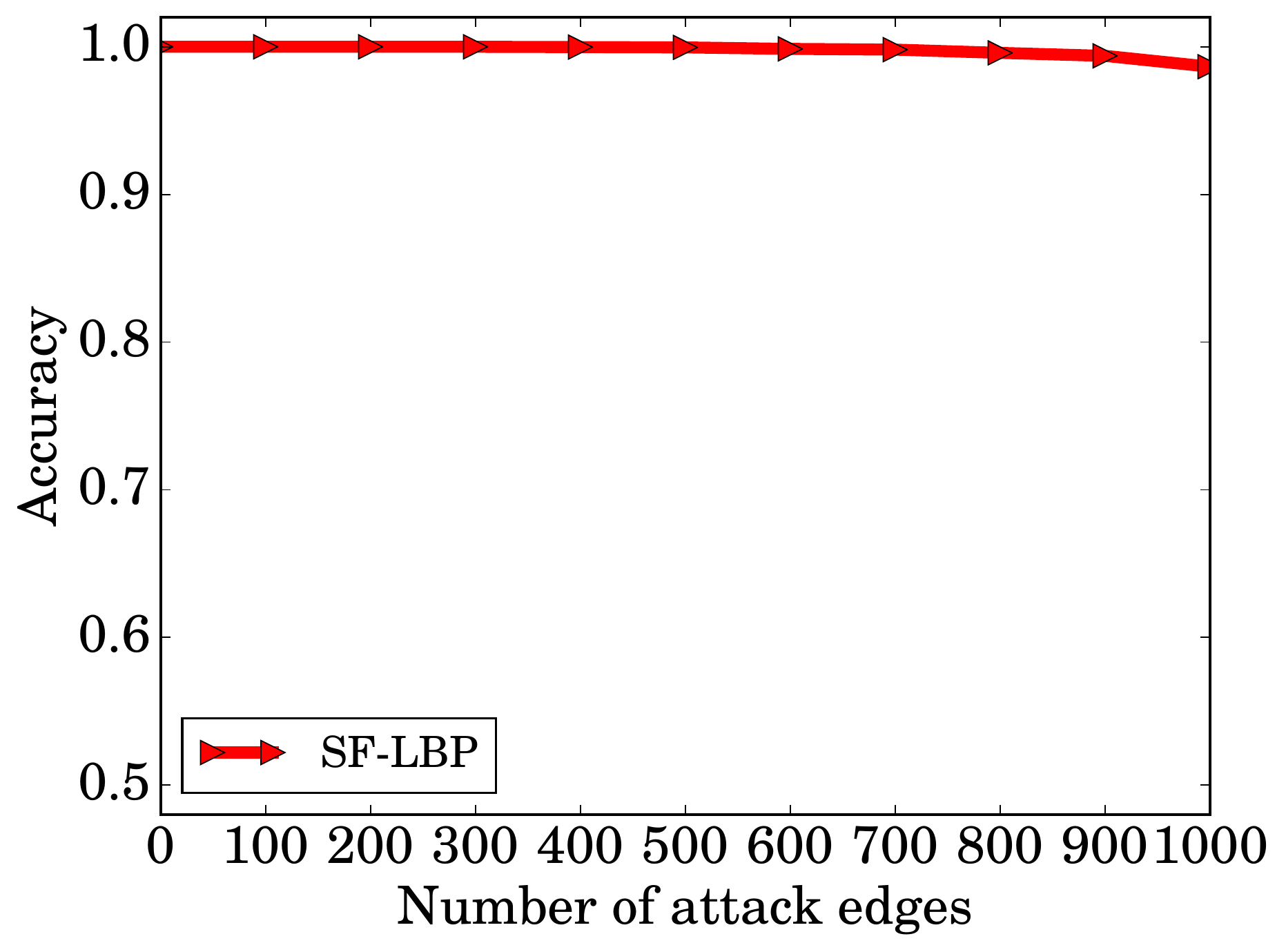}
  \caption{Vary the number of attack edges}
  \label{fig:node_prior_accuracy_2}
\end{subfigure}%
\hspace{0.5cm}
\begin{subfigure}[H]{0.27\textwidth}
  \includegraphics[width=\linewidth]{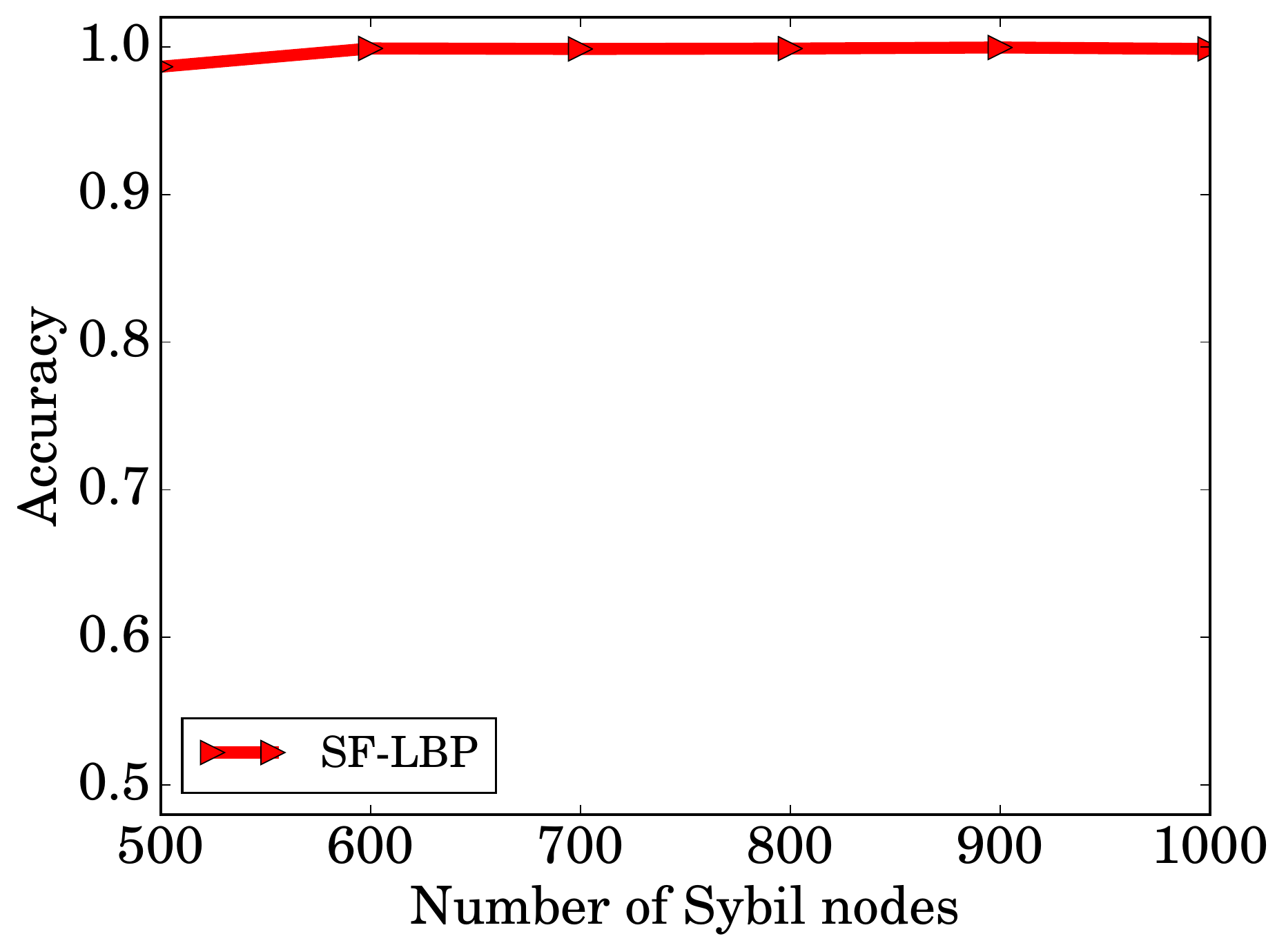}
  \caption{Vary the number of Sybil nodes}
  \label{fig:node_prior_accuracy_3}
\end{subfigure}

\begin{subfigure}[H]{0.27\textwidth}
  \includegraphics[width=\linewidth]{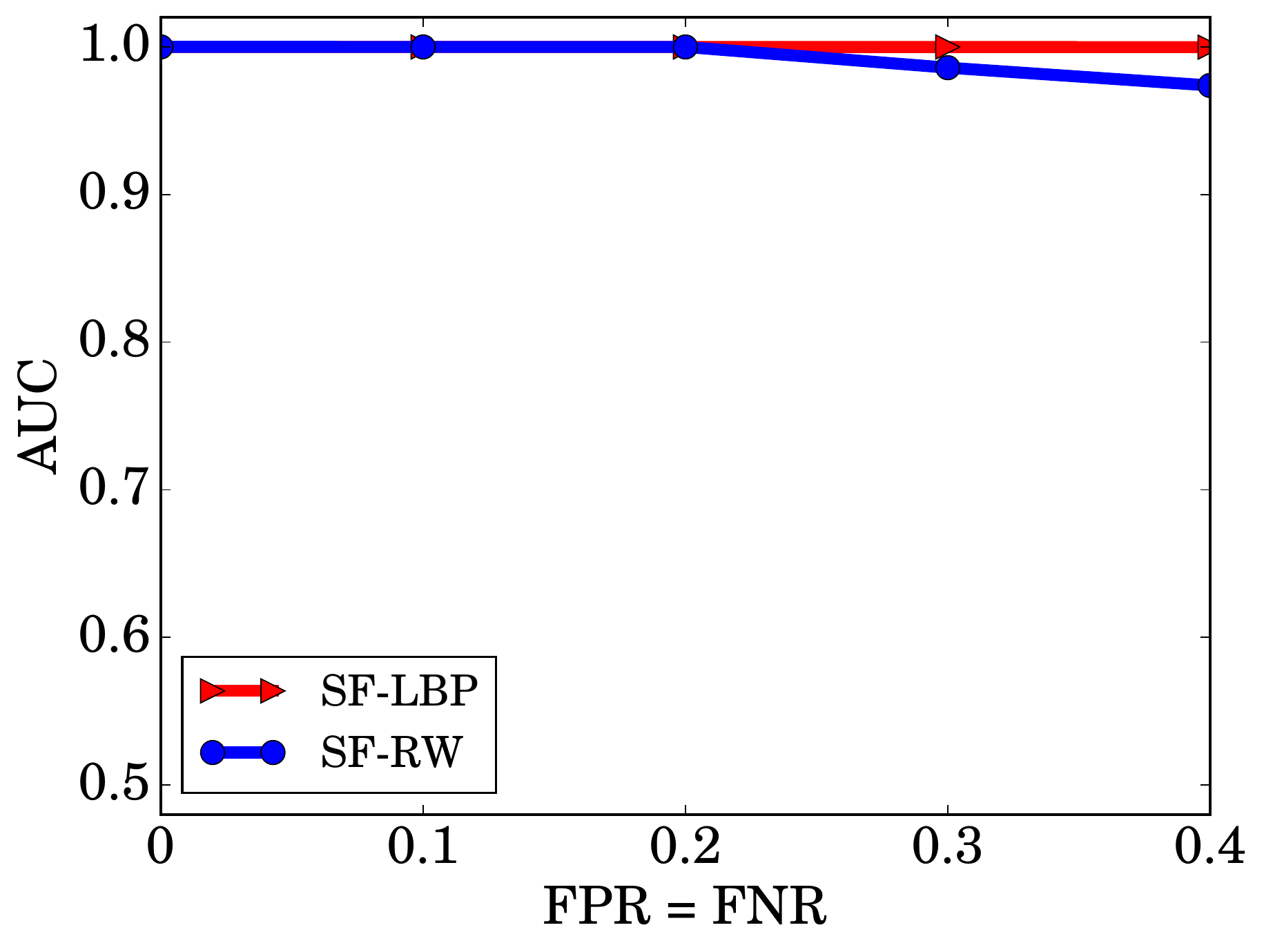}
  \caption{Vary FPR=FNR}
  \label{fig:node_prior_auc_1}
\end{subfigure}%
\hspace{0.5cm}
\begin{subfigure}[H]{0.27\textwidth}
  \includegraphics[width=\linewidth]{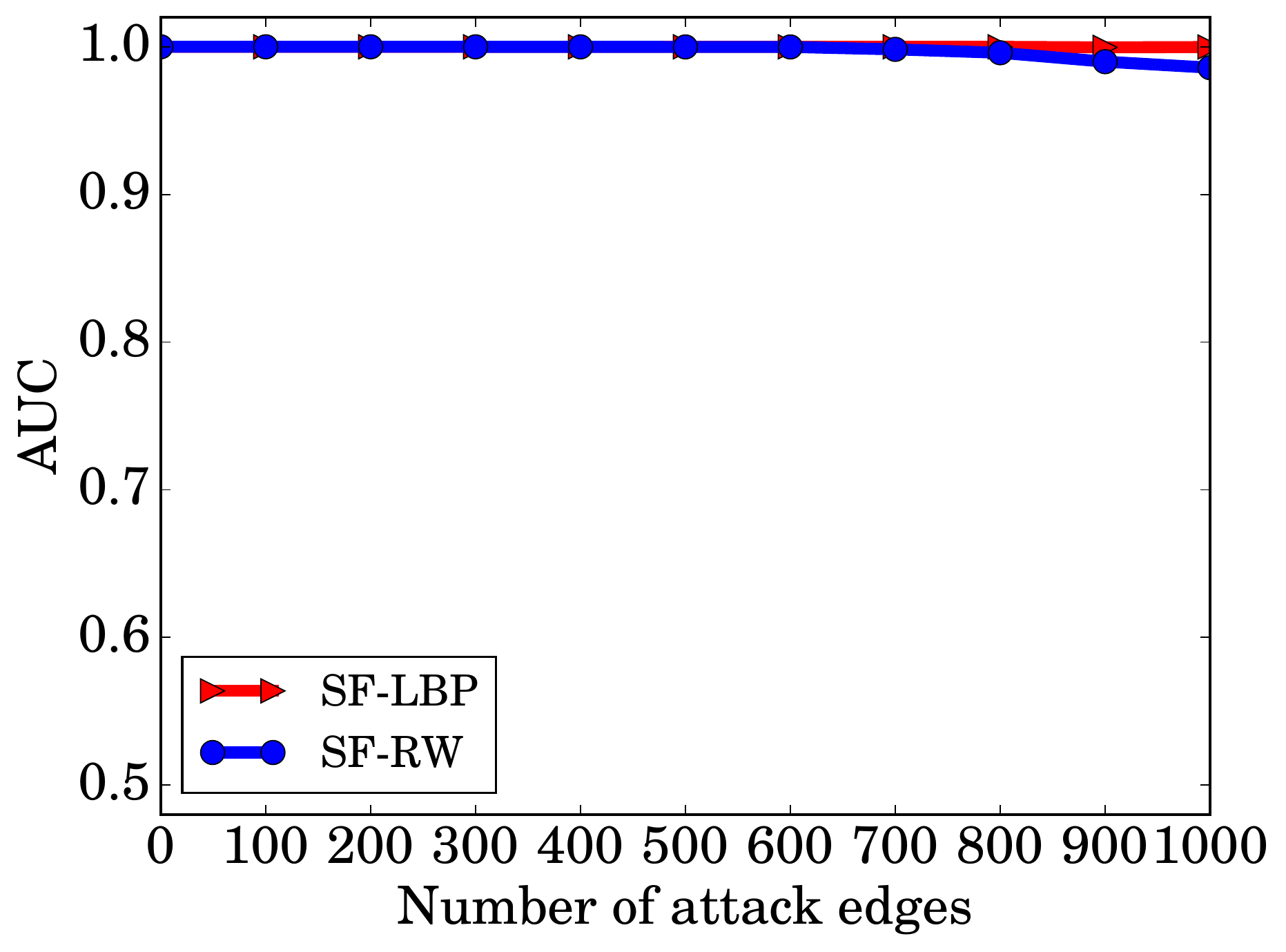}
  \caption{Vary the number of attack edges}
  \label{fig:node_prior_auc_2}
\end{subfigure}%
\hspace{0.5cm}
\begin{subfigure}[H]{0.27\textwidth}
  \includegraphics[width=\linewidth]{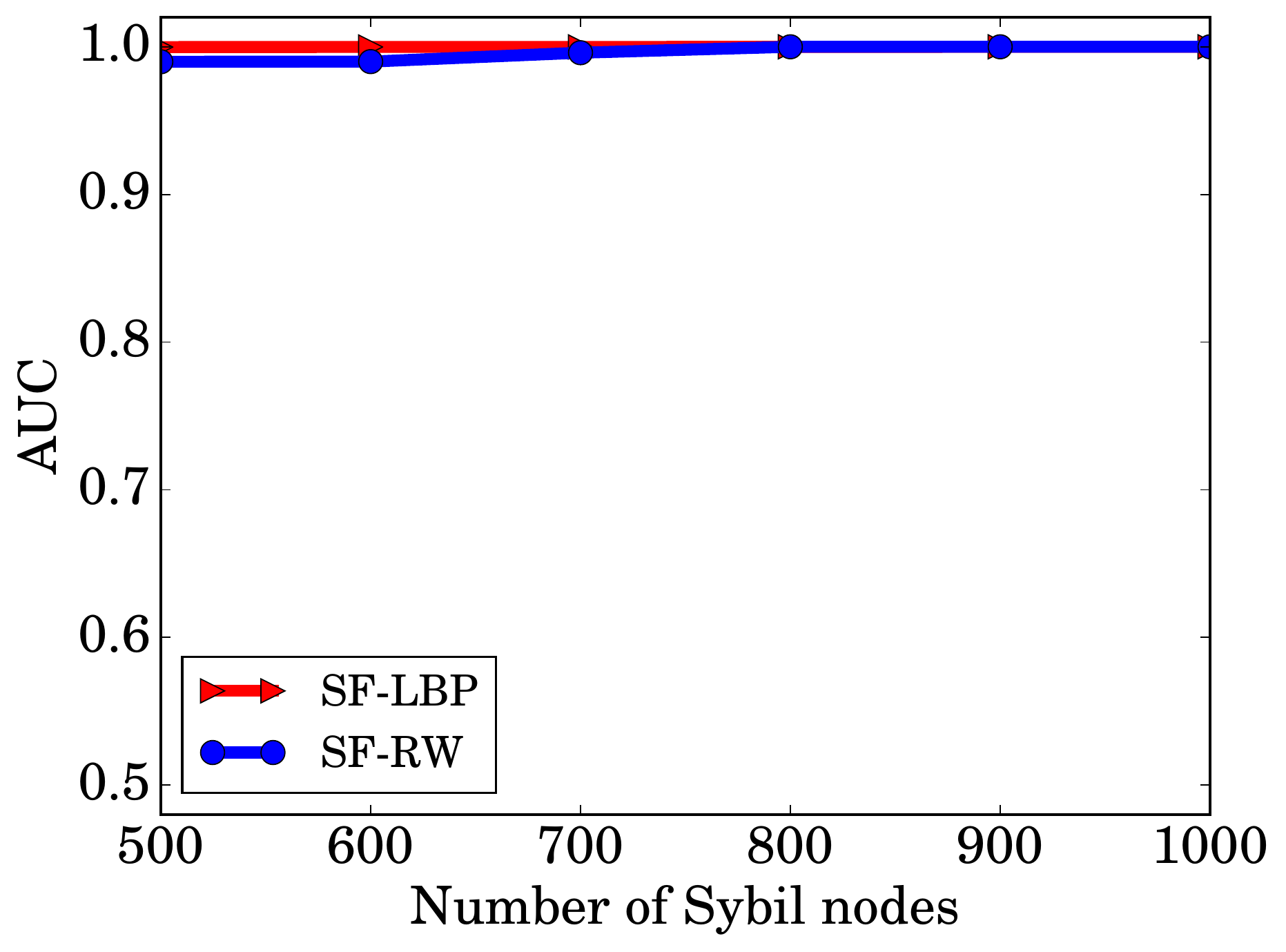}
  \caption{Vary the number of Sybil nodes}
  \label{fig:node_prior_auc_3}
\end{subfigure}
\vspace*{-2ex}
\caption{Accuracy and AUC of \name with weighted random walk (\SFRW{}) and weighted loopy belief propagation (\SFLBP{}) given local \textbf{node} trust scores.
For (a), we vary FPR=FNR of local node classifiers. For (b) and (c), we fix FPR=FNR=0.3.}
\label{fig:node_prior}
\vspace*{-2ex}
\end{figure*}

\begin{figure*}[!h]
\center
\begin{subfigure}[H]{0.27\textwidth}
  \includegraphics[width=\linewidth]{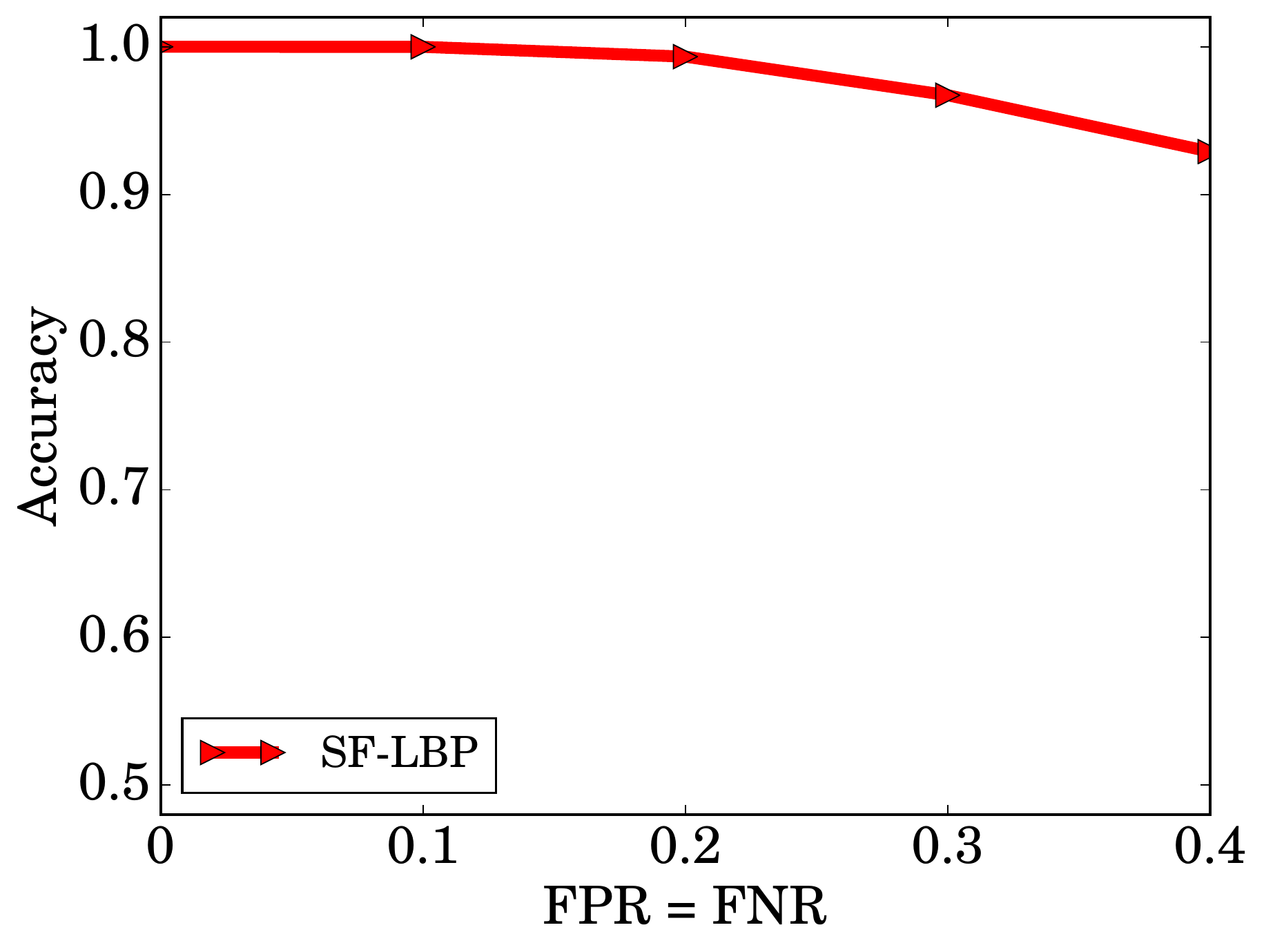}
  \caption{Vary FPR=FNR}
  \label{fig:edge_prior_accuracy_1}
\end{subfigure}%
\hspace{0.5cm}
\begin{subfigure}[H]{0.27\textwidth}
  \includegraphics[width=\linewidth]{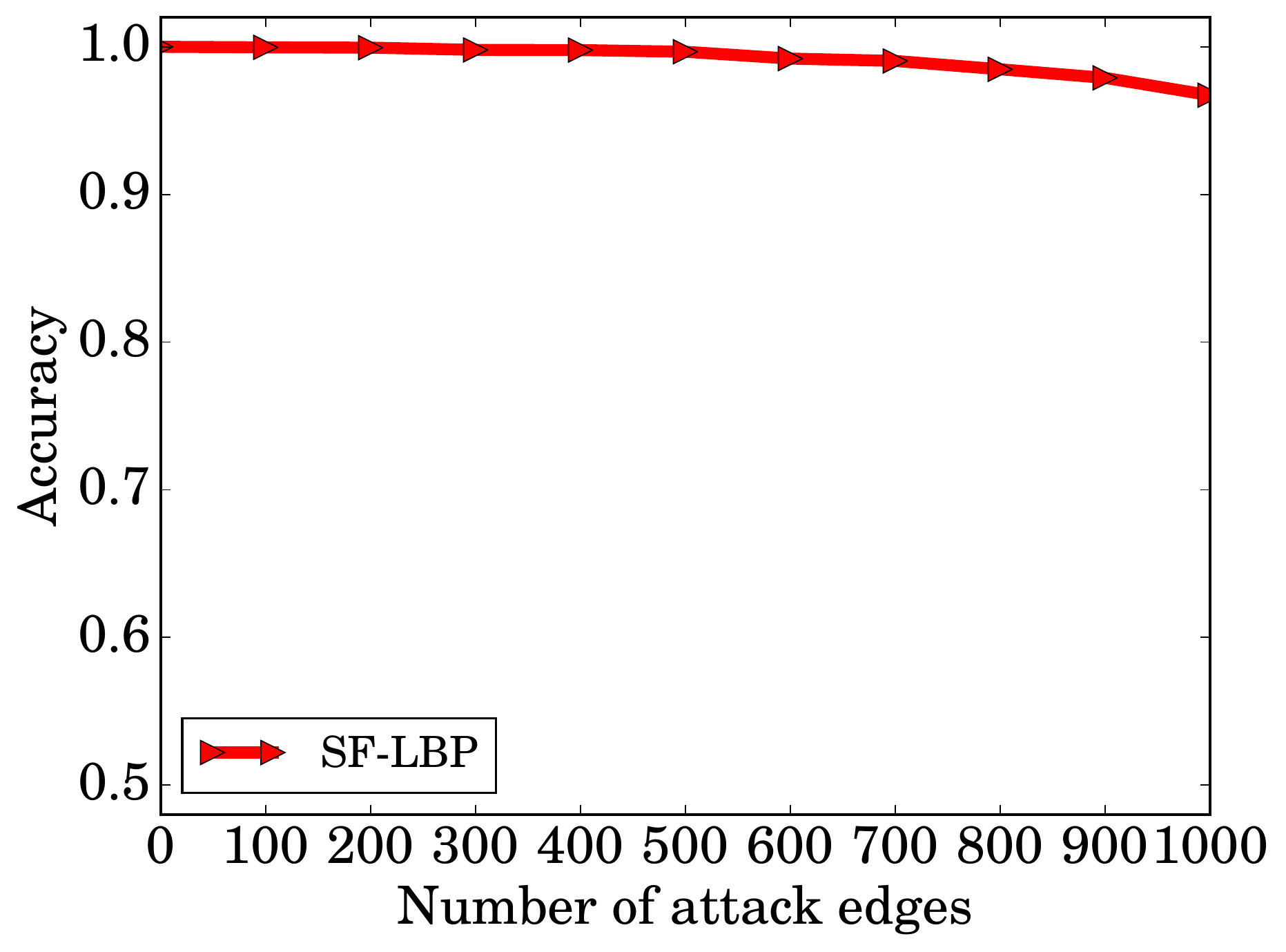}
  \caption{Vary the number of attack edges}
  \label{fig:edge_prior_accuracy_2}
\end{subfigure}%
\hspace{0.5cm}
\begin{subfigure}[H]{0.27\textwidth}
  \includegraphics[width=\linewidth]{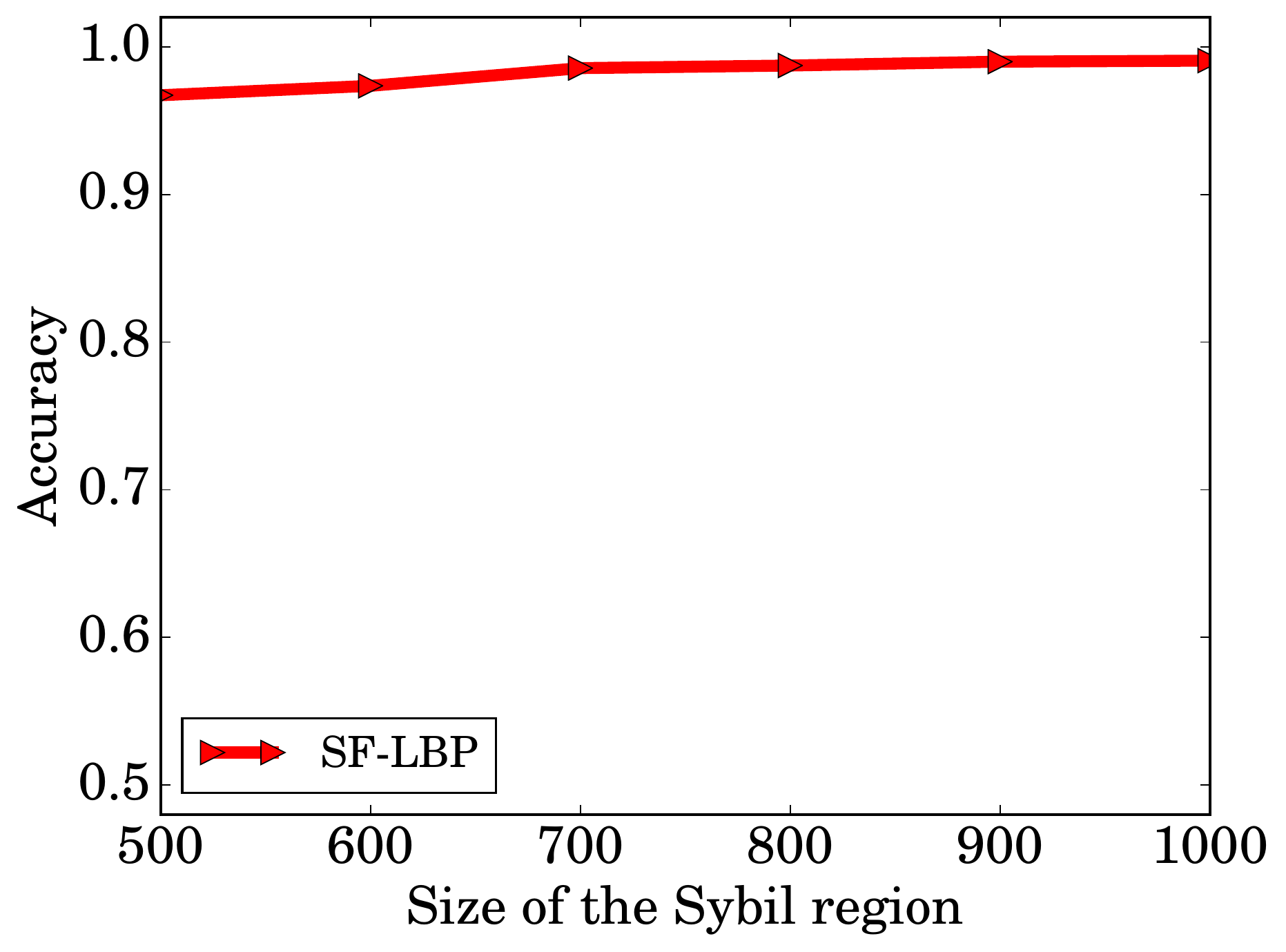}
  \caption{Vary the number of Sybil nodes}
  \label{fig:edge_prior_accuracy_3}
\end{subfigure}

\begin{subfigure}[H]{0.27\textwidth}
  \includegraphics[width=\linewidth]{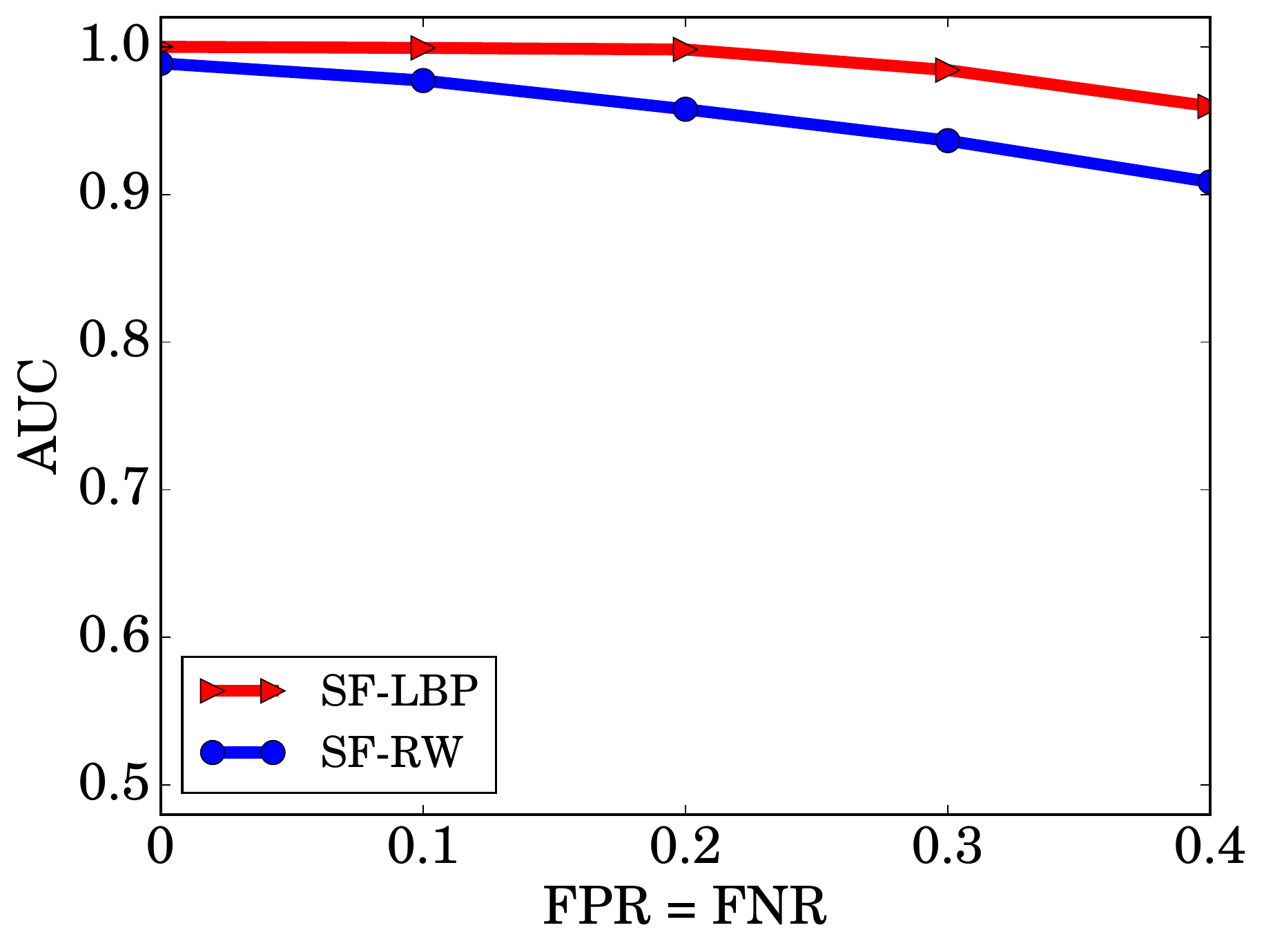}
  \caption{Vary FPR=FNR}
  \label{fig:edge_prior_auc_1}
\end{subfigure}%
\hspace{0.5cm}
\begin{subfigure}[H]{0.27\textwidth}
  \includegraphics[width=\linewidth]{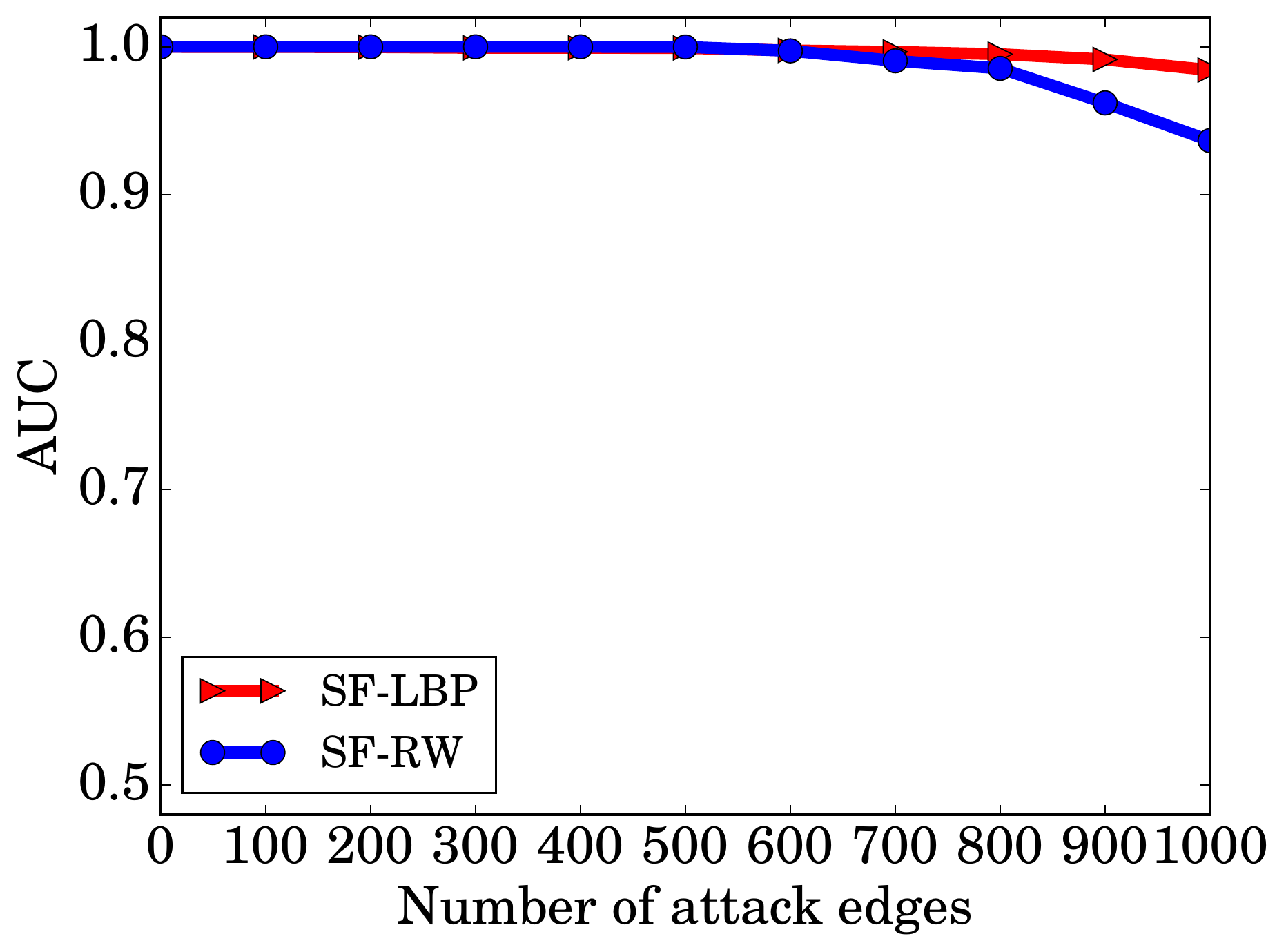}
  \caption{Vary the number of attack edges}
  \label{fig:edge_prior_auc_2}
\end{subfigure}%
\hspace{0.5cm}
\begin{subfigure}[H]{0.27\textwidth}
  \includegraphics[width=\linewidth]{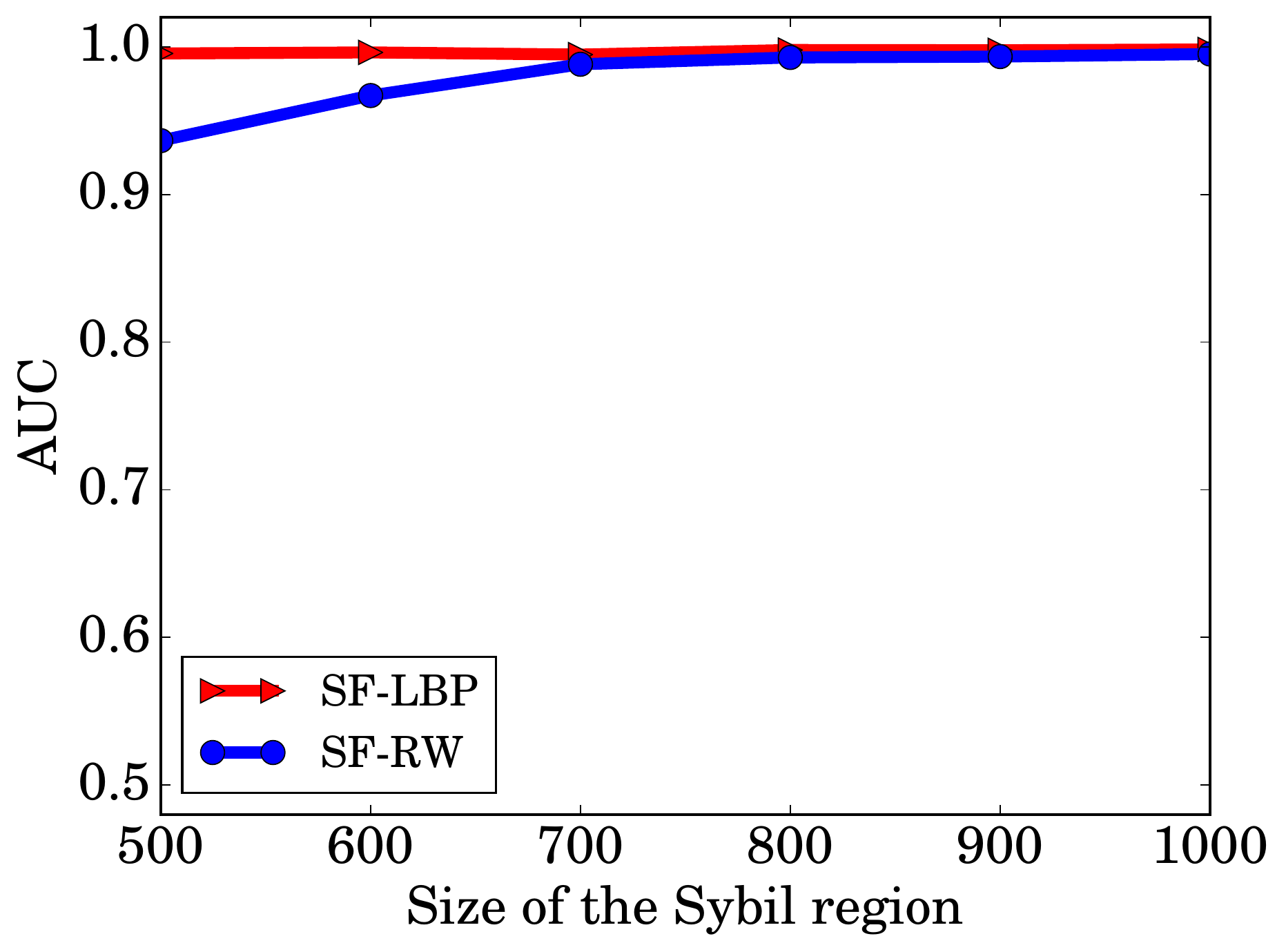}
  \caption{Vary the number of Sybil nodes}
  \label{fig:edge_prior_auc_3}
\end{subfigure}
\vspace*{-2ex}
\caption{Accuracy and AUC of \name with weighted random walk (\SFRW{}) and weighted loopy belief propagation (\SFLBP{}) given local \textbf{edge} trust scores.
For (a), we vary FPR=FNR of local edge classifiers. For (b) and (c), we fix FPR=FNR=0.3.}
\label{fig:edge_prior}
\vspace*{-5ex}
\end{figure*}

\section{Robustness Evaluation}
\label{sec:evaluation_synthetic}

We evaluate the robustness of \name against different network structures
and different levels of local classifier errors.
\subsection{Evaluation Setup}
\label{subsec:sync_setup}

\myparatight{Network generation}
We adopt a similar experimental setting as~\cite{Gong13}, in which we use Preferential Attachment~\cite{Barabasi99} model to generate both the benign region and Sybil region, and randomly add attack edges between the two regions.
In the basic setup, the network consists of 1,000 benign nodes and 500 Sybil nodes. The average degree of the two regions is 10. The number of attack edges is 1,000.

We simulate local trust scores from local classifiers by taking random samples in $[0.1, 0.9]$ with certain false positive rates and false negative rates (0.5 as threshold).
We study the performance of \name under three factors: (1) different levels of errors in local classifiers,
(2) different number of attack edges, and (3) different number of Sybil nodes. When we study one factor, we fix the other factors to be the same as the basic setup and only vary the studied one.

\myparatight{Evaluation metrics}
We measure the accuracy
and the Area Under the Receiver Operating Characteristic (ROC) Curve (AUC)~\cite{auc}
of \name with weighted random walk (\SFRW{}) and \name with weighted LBP (\SFLBP{}). 
Given the ranking of scores of all nodes from the smallest to the largest, AUC measures the the probability that a randomly selected Sybil nodes ranks higher than a randomly selected benign node. A higher AUC indicates better Sybil ranking performance.
For \SFLBP{}, we use a natural threshold 0.5 to compute accuracy. For \SFRW{}, we do not compute accuracy since the final scores of random walk-based approaches (e.g., \cite{sybilrank,Yang12-spam}) are typically very low due to the nature of score distribution
and finding a proper threshold is hard.
\subsection{Evaluation Results}
\label{subsec:sync_res}

Fig.~\ref{fig:node_prior} shows the accuracy and AUC of \name with local node trust scores. We set the edge trust scores to be the default value 0.9 to model homophily. 
For (a), we vary the FPR and FNR of local node classifier from 0 to 0.4 (i.e., from perfect local classifier to better than random guess). For (b) and (c), we set FPR and FNR to be 0.3 (i.e., noisy local classifier) and vary the number of attack edges \& the number of Sybil nodes.
Fig.~\ref{fig:edge_prior} shows the accuracy and AUC of \name with local edge trust scores. We randomly sample 1 benign node and 1 Sybil node as trusted seeds, and set their node trust scores to be 0.9 and 0.1, correspondingly. For the rest of nodes, we set their scores to be the default value 0.5. 

We observe that: (1) when $FPR = FNR \leq 0.3$, \SFRW{} and \SFLBP{} achieve $> 0.98$ accuracy and AUC under all evaluated network settings given local node trust scores, and achieve $>0.92$ accuracy and AUC given local edge trust scores; (2) the performance of both \SFRW{} and \SFLBP{} improves under more accurate local classifiers, less number of attack edges, and more number of Sybil nodes; (3) \SFLBP{} performs better than \SFRW{} in all settings.

\begin{figure*}[!t]
\center
\begin{subfigure}[H]{0.4\textwidth}
  \includegraphics[width=\linewidth]{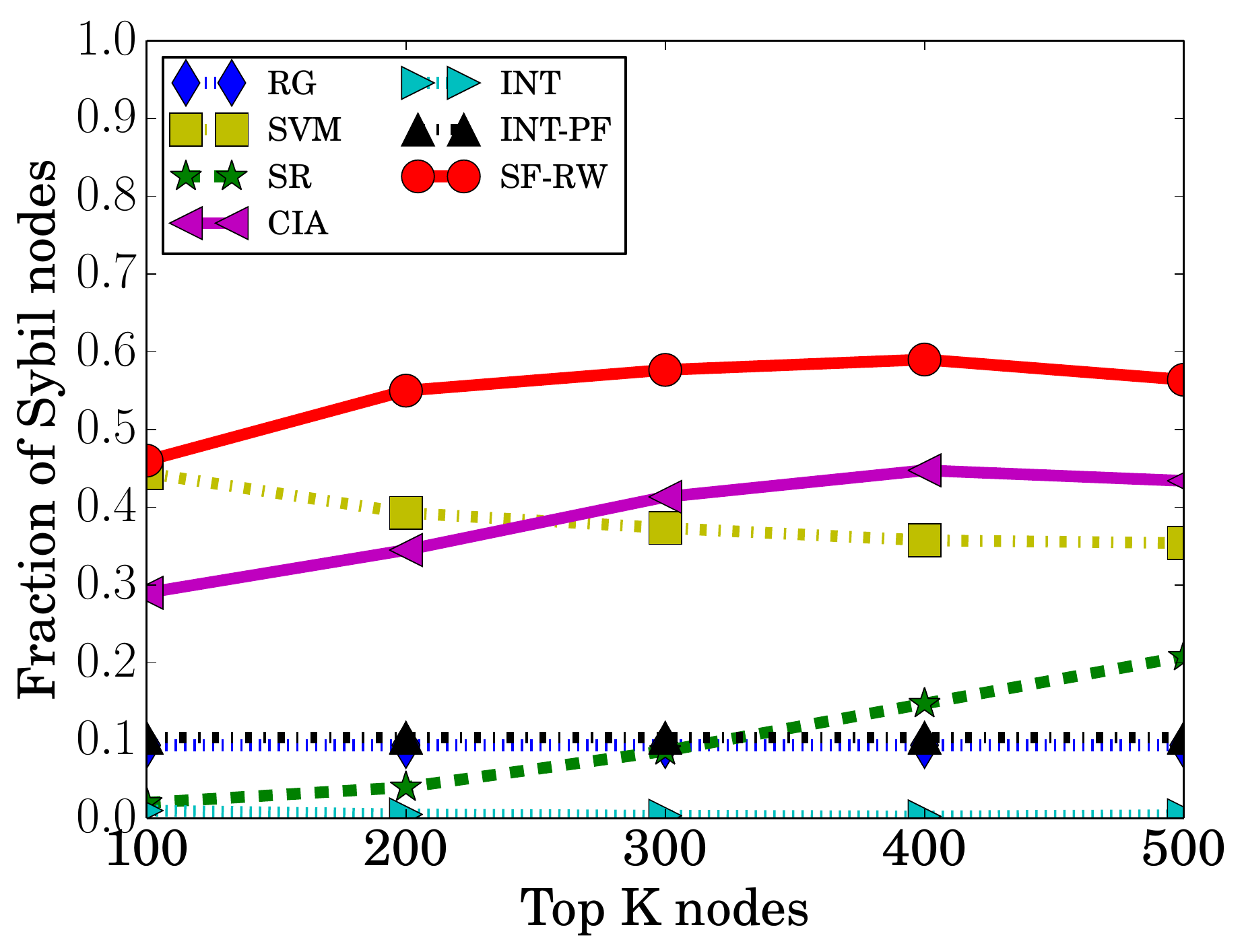}
  \caption{Random walk-based approaches}
  \label{fig:random_walk}
\end{subfigure}%
\hspace{1.75cm}
\begin{subfigure}[H]{0.4\textwidth}
  \includegraphics[width=\linewidth]{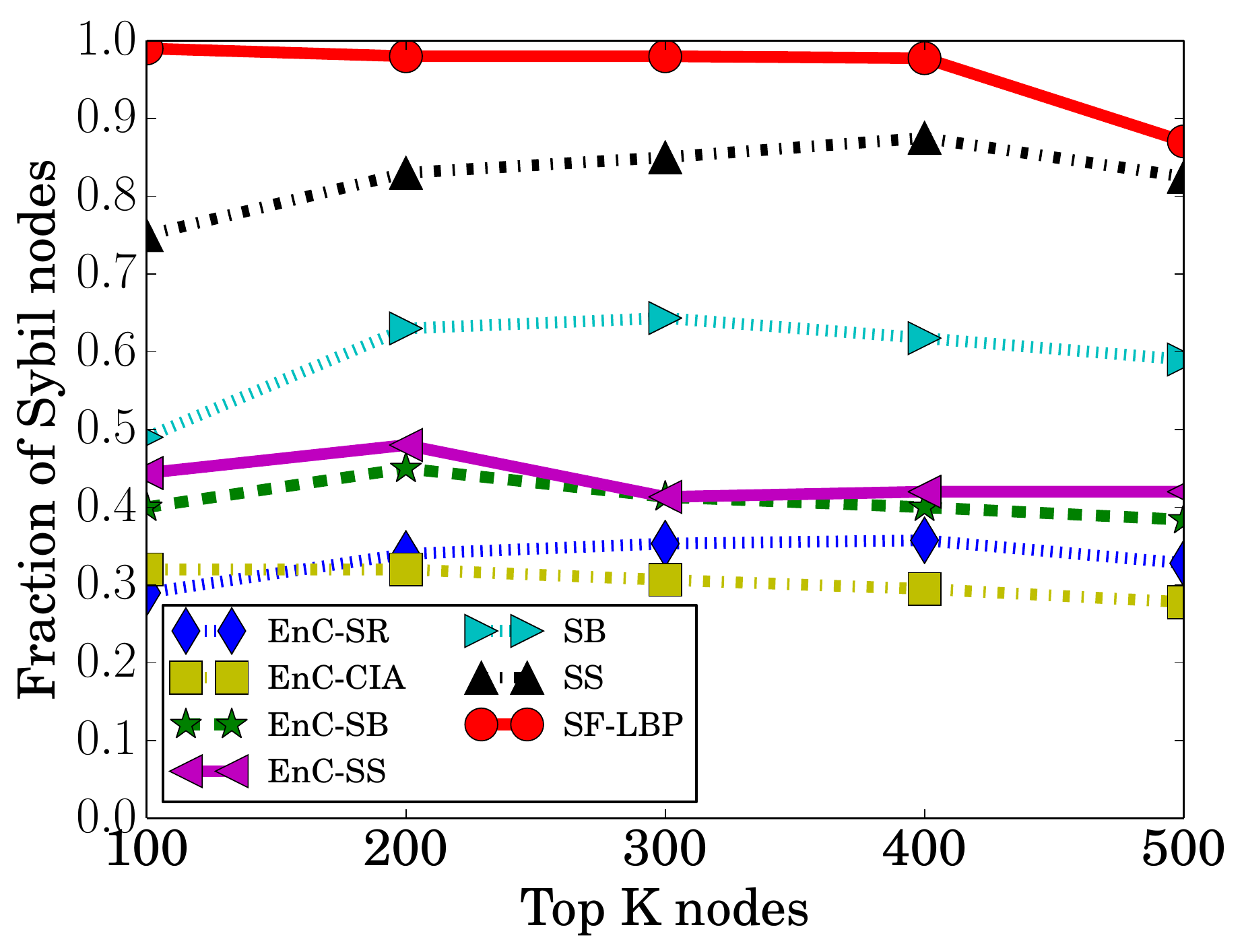}
  \caption{LBP-based approaches and ensemble methods}
  \label{fig:LBP}
\end{subfigure}
\vspace{-3ex}
\caption{
Fraction of Sybils among top K nodes of all evaluated approaches.
We observe that \SFLBP{} has the best ranking performance among all approaches and achieves $> 98\%$ Sybil ranking up to top 400 nodes.
}
\label{fig:small_twitter_ranking}
\vspace{-4ex}
\end{figure*}

\section{Labeled Twitter Evaluation}
\label{sec:small_twitter}

We evaluate \name against state-of-the-art approaches in a real-world, labeled Twitter network. We demonstrate that by combining local attributes with global structure, \name outperforms state-of-the-art significantly.

\subsection{Network Preprocessing and Measurement}
\label{subsec:small_twitter_data}

The original directed network was obtained from~\cite{Yang12-spam}. Since it is easy for the attacker to manipulate one-way directed edges, we followed the established convention~\cite{sybilrank,Yang12-spam,Gong13,wang2017sybilscar,integro} and preprocessed it to an undirected network by retaining an undirected edge if both directions exist. 

After preprocessing, the network consists of 8,167 nodes and 54,146 edges, with verified 7,358 benign nodes and 809 Sybil nodes. 
We observe that: 
(1) Sybil nodes emit a large number of attack edges (40,010), with 49.5 attack edges on average per Sybil; 
(2) 53.4\% of Sybils are isolated (i.e., they only connect to benign nodes). These isolated Sybils emit 37.0\% of attack edges. In addition, 41.2\% of Sybils form a largest connected component (emitting 42.8\% of attack edges), and 5.4\% of Sybils form several small connected groups (emitting 20.2\% of attack edges);
(3) the number of victims is large (5,546; 75.4\% of benign nodes).
Thus, the benign region and the Sybil region can hardly be viewed as separate communities, and the assumptions that existing approaches require (Section~\ref{subsec:approaches}) are not satisfied. As we will show in Section~\ref{subsec:small_twitter_results}, existing approaches~\cite{sybilrank,Yang12-spam,Gong13,wang2017sybilscar,integro} have limited performance on this network.

\subsection{Local Trust Score Computation}
\label{subsec:small_twitter_score_comp}

\myparatight{Feature extraction}
We extract three discriminative node features from the original directed network, and map these features to the corresponding nodes in the undirected network. Since extracting discriminative edge features from this dataset is difficult, we set local edge scores to be 0.9 by default to model homophily.

\begin{itemize}[listparindent=\parindent, leftmargin=*]
\item Incoming requests accepted ratio: $Req_{in}(v) = \frac{\left| In(v) \cap Out(v) \right|}{\left| In(v)\right|}$, where $In(v)$ ($Out(v)$) represents the set of all incoming (outgoing) edges of $v$. 
The insight is that Sybils are more likely to accept incoming requests than benign users in order to quickly propagate spam, resulting in a higher $Req_{in}$. Since the dataset only contains structural information, we use in-degree \& out-degree to model the ratio.

\item Outgoing requests accepted ratio: $Req_{out}(v) = \frac{\left| In(v) \cap Out(v) \right|}{\left| Out(v)\right|}$. 
The insight is that Sybils are less reliable than benign users and hence their outgoing friend requests are less likely to be accepted, resulting in a lower $Req_{out}$.

\item Local clustering coefficient: $CC(v) = \frac{\left| \{  (i, j): i, j\in Nei(v), (i, j)\in E\} \right|}{\left| Nei(v)\right|(\left| Nei(v)\right| - 1)}$), where $Nei(v)$ represents the set of neighbors of $v$.
The insight is that benign users often have well-connected social cliques, and users in such cliques are often friends, resulting in to a high $CC$. 
\end{itemize}

\myparatight{Training a SVM classifier}
We randomly sample 50 benign nodes and 50 Sybil nodes as the training set, and train a SVM classifier with RBF kernel using \emph{LIBSVM}~\cite{Chang2011}. We then output probabilistic estimates as local node scores. 
Note that we extract these features and adopt the SVM classifier in particular for this Twitter network. The system administrator is free to explore other features and classifiers under the overall \name framework.

\subsection{Evaluated Approaches} 
\label{subsec:small_twitter_approaches}
For \name, we propagate local scores through 
weighted random walk (\SFRW{}) and weighted loopy belief propagation (\SFLBP{}).
We also evaluate the following existing approaches: 
(1) local node classification: \SVM{}; 
(2) global structure-based approaches: SybilRank (\SR{}), CIA (\CIA{}), SybilBelief (\SB{}), SybilSCAR (\SSC{}), \'{I}ntegro (\INTEGRO{}), \'{I}ntegro with perfect victim prediction (\INTEGROPF{}). We use the same training data as propagation seeds. For \INTEGRO{}, we further sample 50 victims and 50 non-victims to learn a victim predictor based on Random Forest algorithm~\cite{integro}. For \INTEGROPF{}, we model a perfect victim predictor by setting the probability score of each victim to be 1 and the score of each non-victim to be 0;
(3) ensemble approaches: \ENCSR{}, \ENCCIA{}, \ENCSB{}, \ENCSS{}. We combine the scores from SVM classifier with the scores from structure-based approaches in a standard voting scheme;
(4) random guess: \RG{}.

\begin{figure*}[t]
	\centering
	\includegraphics[width=0.75\textwidth]{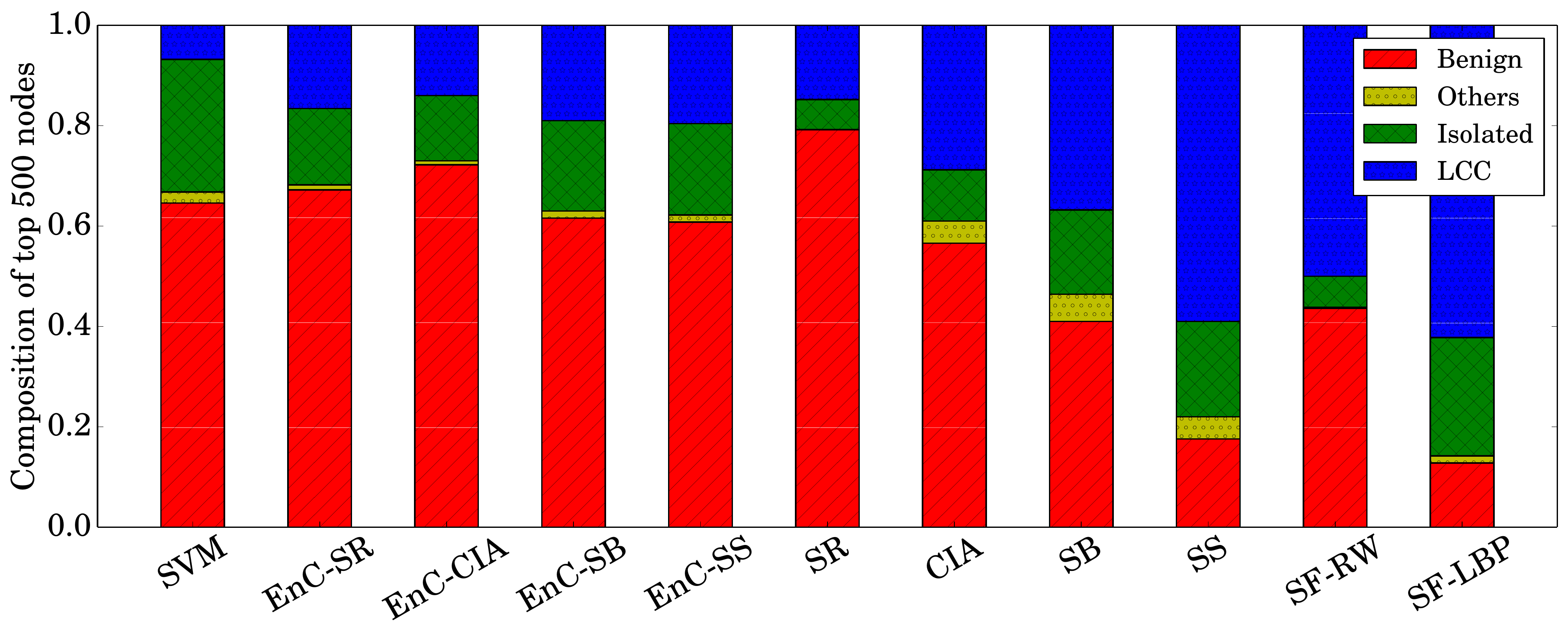}
	\caption{Decomposition of top 500 nodes. We observe that \SFLBP{} has the best performance in ranking both isolated Sybils and Sybils in the largest connected component (LCC).}
	\label{fig:small_twitter_composition}
	\vspace{-5ex}
\end{figure*}
\subsection{Evaluation Results}
\label{subsec:small_twitter_results}

\myparatight{Sybil ranking performance}
Following the established convention~\cite{sybilrank,Gong13,integro,wang2017sybilscar}, we evaluate the \emph{ranking performance} of the compared approaches by ranking all nodes in an ascending order of their final scores. A more effective approach will rank more Sybil nodes upfront.

Fig.~\ref{fig:small_twitter_ranking} shows the fraction of Sybils among top K nodes.
We observe that:
(1) \SFRW{} has the best performance among all random walk-based approaches. Specifically, the average improvement of \SFRW{} w.r.t. \SR{}, \CIA{}, and \INTEGRO{} is 44.8\%, 16.2\%, and 54.3\%, respectively (the improvement of $A$ w.r.t. $B$ is computed as $A-B$, where $A$ and $B$ represent the fraction of Sybils);
(2) \SFLBP{} has the best performance among all evaluated approaches. Specifically, the average improvement of \SFLBP{} w.r.t. \RG{}, \SVM{}, \SR{}, \CIA{}, \INTEGRO{}, \SB{}, and \SSC{} is 85.7\%, 57.5\%, 85.9\%, 57.4\%, 95.5\%, 36.5\%, and 13.4\%, respectively. Besides, \SFLBP{} achieves $> 98\%$ Sybil ranking up to top 400 nodes;
(3) \SFRW{} and \SFLBP{} outperform \INTEGRO{}, \INTEGROPF{}, and ensemble approaches significantly. 
Surprisingly, even under perfect victim prediction, the performance of \INTEGROPF{} is just slightly better than \RG{}. In fact, \'{I}ntegro assigns low weights to all edges originated from victims according to the formula $w(v_i, v_j ) = \min \{1, \beta\cdot (1 - \max\{p(v_i), p(v_j )\})\}$~\cite{integro}. Hence, the propagation of scores from trusted seeds to victims and from victims to other benign nodes is inhibited. When the number of victims is large (75.4\% in our Twitter network), even under perfect victim prediction, a large number of benign nodes will receive low final scores, including the victims and the benign nodes that are separated from propagation seeds by victims. This problem becomes worse in practice since it is impossible to have perfect victim prediction. For \INTEGRO{}, the victim predictor has 18.3\% precision and 19.6\% recall, and we can see that \INTEGRO{} has nearly 0 Sybil ranking capability. In short, \INTEGRO{} has very limited performance when the number of victims is large or the victim predictor is inaccurate.

\myparatight{Measuring the composition of top-ranked Sybils}
We study the power of each approach in ranking different types of Sybils: isolated Sybils (\textit{Isolated}), Sybils in the largest connected component (\textit{LCC}), and Sybils in small, connected groups (\textit{Others}).
Fig.~\ref{fig:small_twitter_composition} shows the decomposition of top 500 nodes. We omit \INTEGRO{} and \INTEGROPF{} due to their very limited performance.
We observe that:
(1) local node classification approach (\SVM{}) is more powerful in ranking isolated Sybils (74.6\% of all Sybils ranked by \SVM{});
(2) global structure-based approaches are more powerful in ranking Sybils in the LCC (71.2\%, 66.4\%, 62.4\%, and 71.6\% of all Sybils ranked by \SR{}, \CIA{}, \SB{}, and \SSC{}, respectively);
(3) \SFLBP{} has the best performance in ranking both isolated Sybils and Sybils in the LCC.

\myparatight{Summary}
In summary, \name methods (\SFRW{}, \SFLBP{}) significantly outperform existing approaches in terms of Sybil ranking, and the most effective approach is to propagate local trust scores from local classifiers through weighted loopy belief propagation (\SFLBP{}).

\section{Large-Scale Labeled Twitter Evaluation}
\label{sec:large_twitter}

We further evaluate \name against state-of-the-art approaches in a real-world, large-scale, labeled Twitter network comprising over 20 million nodes and 265 million edges. 

\subsection{Ground Truth Collection}
\label{subsec:large_twitter_data_col}

We obtained a snapshot of the Twitter follower network from~\cite{Kwak10}. Similar to Section~\ref{subsec:small_twitter_data}, we preprocessed the original directed network to an undirected network by only retaining bi-directional edges. After preprocessing, the network consists of 21,297,772 nodes and 265,025,545 edges. To collect ground truth, we re-crawled every account using Twitter's API, which told us the account status. We found that 145,156 (0.7\%) nodes were suspended by Twitter, 1,911,482 (9.0\%) nodes were deleted, and the rest were still active. We treated the suspended accounts as Sybil nodes and the active accounts as benign nodes. For the deleted accounts, since we were not sure whether they were deleted by users or by Twitter, we did not include them in the training and evaluation.

\subsection{Network Structure Measurement}
\label{subsec:large_twitter_measurement}

We adopt \emph{modularity}~\cite{Newman04}, ranging from -0.5 to 1, to quantify if a network partition can be viewed as two communities. Clauset et al.~\cite{Clauset04} concluded that modularity $> 0.3$ indicates a significant community structure. However, we find that the partition consisting of the benign and Sybil regions only has modularity 0.0042. Thus, the two regions cannot be viewed as separate communities.
Next, we show two reasons:

\emph{(1) Half of Sybils are isolated:} In total, we find 77,917 connected components in the Sybil region. Among them, 50\% of Sybils are isolated, 45\% of Sybils form a largest connected component, and the rest of Sybils form small connected components whose sizes are less than 20. 
Furthermore, we find that the modularity of the partition consisting of the benign region and the largest Sybil connected component is still only 0.0046, which means that even this largest connected component cannot be viewed as a community.

\emph{(2) Large number of attack edges:} In total, there are 18,414,469 attack edges, which means each Sybil node successfully attacks around 127 benign nodes on average. 
Furthermore, we find that the number of neighbors of both benign and Sybil nodes follow long-tail distributions, which is also widely observed in other OSNs such as LiveJournal~\cite{Mislove07} and Google+~\cite{Gong12-imc}. Thus, we speculate that Sybils are imitating the benign region to evade automatic detection. Besides, 90\% of attack edges concentrate on 3\% of benign nodes. Thus, we speculate that such nodes are celebrities that tend to follow back any account that follows them.

Note that these structural properties of Sybil nodes in our Twitter network match those in another large-scale Twitter network~\cite{Ghosh12} and those in the RenRen network~\cite{Yang11-sybil}, which indicates the representativeness of our observations. 
As a result, the assumptions that existing approaches require are not satisfied, leading to their limited performance (Section~\ref{subsec:large_twitter_results})

\subsection{Local Trust Score Computation}
\label{subsec:twitter_nodeprior}

\begin{figure}[t]
\begin{subfigure}[H]{0.22\textwidth}
  \includegraphics[width=\linewidth]{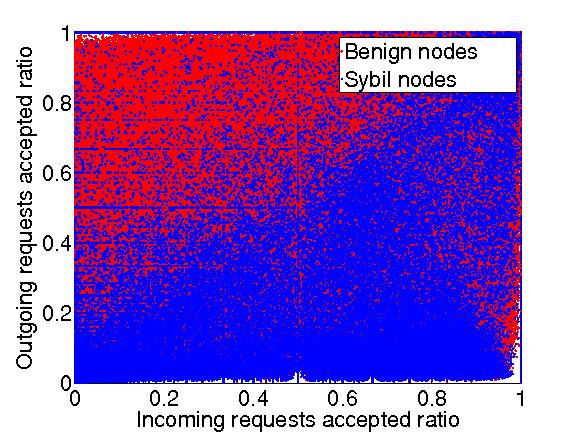}
\caption{Scatter plot}
  \label{fig:twitter_node_feature_scatter}
\end{subfigure}%
\begin{subfigure}[H]{0.22\textwidth}
  \includegraphics[width=\linewidth]{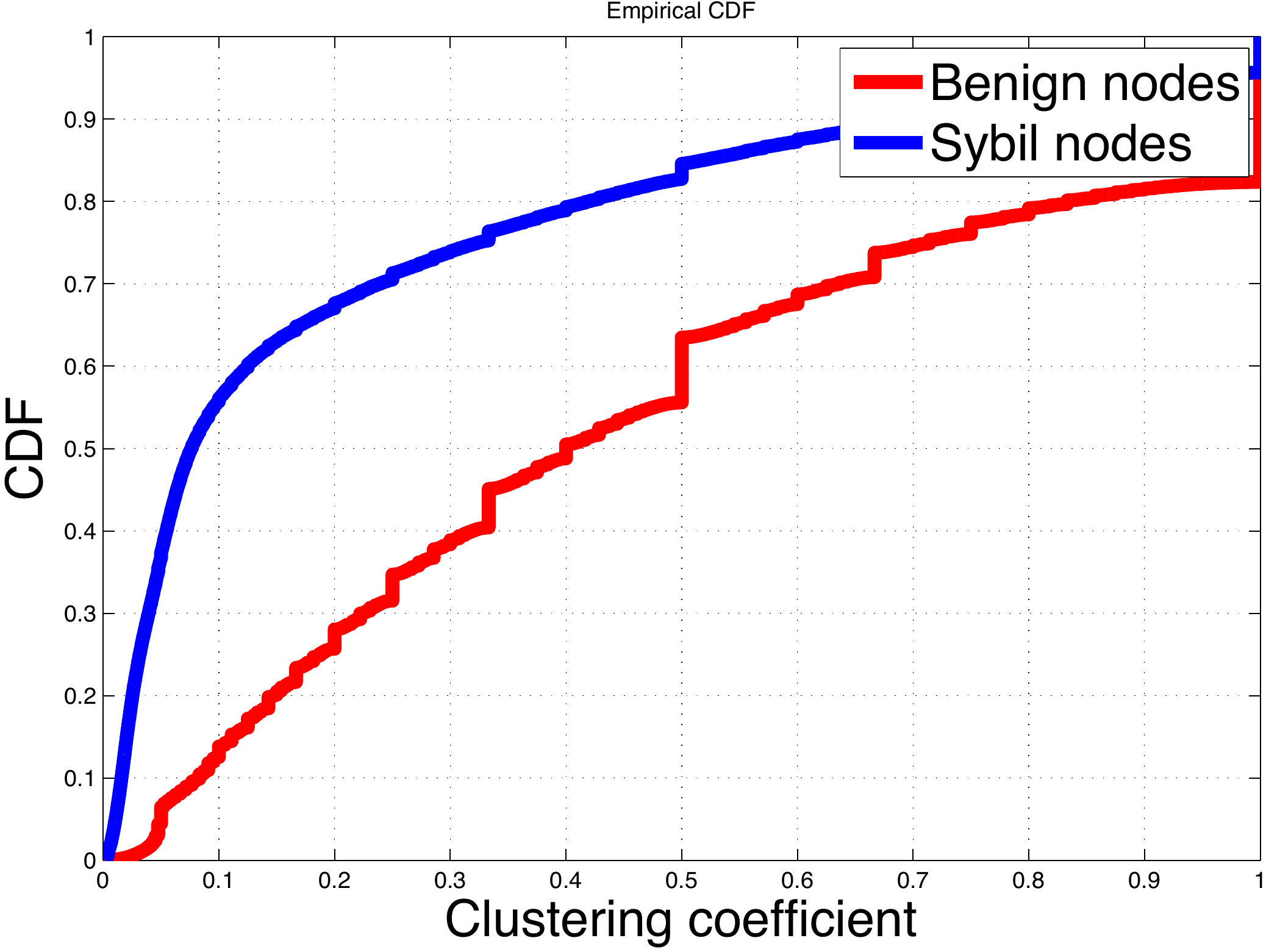}
\caption{CDF}
  \label{fig:twitter_node_feature_cc}
\end{subfigure}%
\vspace{-3ex}
\caption{Analysis of local node features}
\label{fig:twitter_node_feature}
\vspace{-5ex}
\end{figure}

We extract the same set of features as Section~\ref{subsec:small_twitter_score_comp}. 
Fig.~\ref{fig:twitter_node_feature} shows the scatter plot of $Req_{out}$ vs. $Req_{in}$ and the CDF plot of $CC$.
As expected, Sybil nodes tend to have a higher $Req_{in}$, a lower $Req_{out}$, and a lower $CC$.
We randomly sample 3,000 benign nodes and 3,000 Sybil nodes, and train a SVM classifier using \emph{LIBSVM}~\cite{Chang2011}. We then output probabilistic estimates as local node scores. 
Since extracting discriminative edge features from this dataset is difficult, we set local edge scores to be 0.9 by default to model homophily.

\subsection{Evaluation Results}
\label{subsec:large_twitter_results}
Following the notations in Section~\ref{subsec:small_twitter_approaches} and the established convention~\cite{sybilrank,Gong13,integro,wang2017sybilscar}, we evaluate the \emph{ranking performance} of \name approaches (\SFRW{}, \SFLBP{}) against state-of-the-art Sybil defense approaches \SR{}, \CIA{}, \INTEGRO{}, \INTEGROPF{}, \SB{}, and \SSC{} (we follow the same notation as Sec.~\ref{subsec:small_twitter_approaches}). We measure two metrics: (1) AUC; (2) fraction of Sybils among top K ranked nodes.

\begin{table}[t]
\caption{AUC in the large-scale Twitter network}\label{tab:large_twitter_auc}
\centering
\begin{adjustbox}{width=0.42\textwidth}
\begin{tabular}{|c|c|c|c|c|c|c|c|}
\hline
\SR{}	&\CIA{}	&\INTEGRO{}	&\INTEGROPF{}	&\SB{}	&\SSC{}	&\SFRW{}	&\SFLBP{}\\\hline0.57 &0.80 &0.48 &0.54 &0.74 &0.74 &0.81 &0.85\\\hline
\end{tabular}
\end{adjustbox}
\vspace{-4ex}
\end{table}

\begin{figure}[t]
\centering
\includegraphics[width=0.425\textwidth]{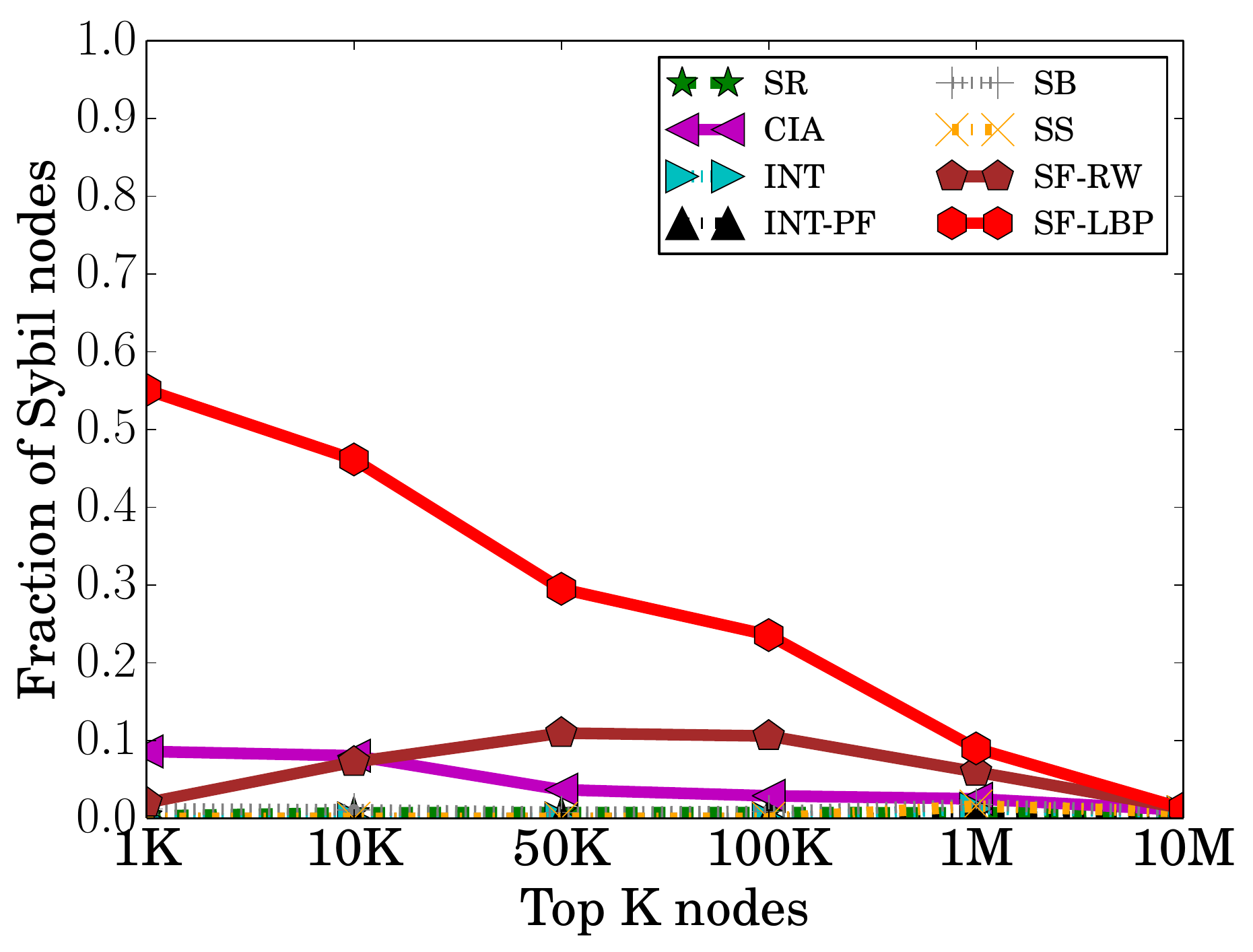}
\caption{Fraction of Sybils in the large-scale Twitter network} 
\label{fig:large_twitter_ranking}
\vspace{-5ex}
\end{figure}

Table~\ref{tab:large_twitter_auc} shows the AUC of the evaluated approaches and Fig.~\ref{fig:large_twitter_ranking} shows the fraction of Sybils among top K nodes.
We observe that:
(1) \SFRW{} has the best AUC among all random walk-based approaches. Specifically, the improvement of \SFRW{} w.r.t. \SR{}, \CIA{}, \INTEGRO{}, and \INTEGROPF{} is 0.24, 0.01, 0.33, and 0.27, respectively (measured as $A-B$);
(2) \SFLBP{} has the best AUC among all evaluated approaches. Specifically, the average improvement of \SFLBP{} w.r.t. \SR{}, \CIA{}, \INTEGRO{}, \INTEGROPF, \SB{}, and \SSC{} is 0.28, 0.05, 0.37, 0.31, 0.11, and 0.11, respectively; 
(3) \SFLBP{} outperforms state-of-the-art Sybil defenses significantly in terms of Sybil ranking. Specifically, among the top 1K nodes, the Sybil ranking improvement of \SFLBP{} w.r.t. \SR{}, \CIA{}, \INTEGRO{}, \INTEGROPF, \SB{}, and \SSC{} is 54.8\%, 46.5\%, 55.1\%, 55.1\%, 55.1\%, 53.9\%, and 55.1\%, respectively.

\subsection{Limitations in Twitter's Sybil Detection Policy} 
\label{subsubsec:top_100}

Recall that we obtained our ground truth based on whether the account was active or suspended by Twitter. Thus, it is possible that some accounts are actually Sybil but evade Twitter's detection policy. To test this, we manually examined the top 100 accounts, of which 71 were suspended and 29 were active. Among the 29 active accounts, we found: (1) 3 benign accounts, with a long timeline and diverse tweets; (2) 24 Sybil accounts, with common low-quality profile pictures and spam tweets;
(3) 2 suspicious accounts with protected information. 
\emph{These 24 (24/29=82.8\%) active Sybil accounts evaded Twitter's detection policy but were successfully captured by \name's ranking mechanism.}

\eat{
\myparatight{Validation from Twitter's security team}
We validated our top 1K list with Twitter's security team. Among a random sample of 278 accounts investigated by Twitter, 200 (71.9\%) accounts were Sybil, 57 accounts were benign, 14 accounts were compromised, and 7 accounts were unknown. Unfortunately, due to the privacy concerns, Twitter did not provide us details of the random sample as well as the evidence used in their investigation. Even though, we re-crawled this top 1K list immediately after our communications with Twitter, and found that 52 additional accounts got suspended. 
Thus, we conclude that: (1) Twitter's Sybil detection policy has limitations and \name is able to capture a large portion of Sybil accounts that Twitter fails to detect; (2) given more accurate ground truth, \name's measured performance could potentially be higher than what we reported.

}

\section{Discussion}
\label{sec:discussion}

\myparatight{Resilience against social churn}
According to~\cite{temporal-sybil}, existing Sybil defenses such as SybilInfer~\cite{Danezis09} and SybilRank~\cite{sybilrank} are vulnerable to the churn in social graphs. To attack these systems, \cite{temporal-sybil} considers churn in attack edges by gradually moving the attack edges closer to the trusted seeds, so that the seed-based score propagation will eventually fail. The success of this temporal attack requires two assumptions: (1) the attacker knows the location of the trusted seeds in the social graph; (2) the location of the trusted seeds will not change drastically for a certain period of time, so that the attacker has enough time to perform the attack.
However, we propose that the system administrator can frequently change the trusted seeds and rerun the detection program. Furthermore, \name does not start propagation from the trusted seeds. Instead, \name computes local scores for all nodes and propagates these scores. As a result, the attacker does not have a direction for moving the attack edges gradually.

\myparatight{Strategic adversaries}
Strategic adversaries who are aware of the features used in \name's local classifiers 
may attempt to evade detection by changing attack strategies. \name's multi-layer protection restricts such possibilities by combining heterogeneous information sources. 
Specifically, if the attacker aims to evade the detection of local node classifier by mimicking the local features of benign users (i.e., lower $Req_{in}$, higher $Req_{out}$, and higher $CC$), he needs to establish more connections between Sybil identities.
Consequently, Sybils will be much more densely connected, and the trust score propagation module (Fig.~\ref{fig:sybilframe}) will be more effective to detect them. Machine learning in these adversarial scenarios remains a challenging problem for the research community. In general, we advocate periodically retraining the local classifiers to respond to the dynamics of attack behaviors.

\myparatight{Broader applicability} Our framework that combines local attributes with global structure has broad applicabilities for network security. For example, the area of botnet detection can benefit from similar techniques that combine host-level information with network structure-based information.

\section{Conclusion and Future Work}
\label{sec:conclusion}

In this work, we proposed \name, a defense-in-depth framework that novelly combines local attributes with global structure to perform robust Sybil detection. \name overcomes the limitations in existing approaches by leveraging local attributes to train local classifiers and propagating the local classifier scores through global structure via weighted score propagation. Experimental results in synthetic topologies and real-world topologies demonstrate that \name outperforms state-of-the-art approaches significantly.
In future, we plan to generalize \name to detect Sybils in \emph{directed} social graphs~\cite{wang2017gang} and 
apply \name to other security applications such as botnet detection and reputation systems.

\section*{Acknowledgement}

This work was supported in part by DARPA RADICS under contract FA-8750-16-C-0054, and the National Science Foundation under grants CNS-1553437 and CNS-1409415.

\bibliographystyle{IEEEtran}
\bibliography{refs}

\end{document}